\documentclass[%
 reprint,
superscriptaddress,
 amsmath,amssymb,
 aps,
]{revtex4-1}

\usepackage[utf8]{inputenc} 
\usepackage{mathtools} 
\usepackage[pdftex]{graphicx}
\usepackage{graphicx}
\usepackage{dcolumn}
\usepackage{bm}
\usepackage{amsmath}
\usepackage{amsfonts}

\usepackage{xr-hyper}
\newcommand{\Lundmed}{Department of Medical Radiation Physics, Lund University, Lund, Sweden}
\newcommand{\LundSR}{Division of Synchrotron Radiation Research, Lund University, Lund, Sweden}
\newcommand{\WISE}{Wallenberg Initiative Materials Science for Sustainability, Lund University, Lund, Sweden}
\newcommand{\Nanolund}{NanoLund, Lund University, Lund, Sweden}
\begin{document}

\title{Breaking order: Talbot effect with spinodal architectures}

\author{Robin Kr\"uger}
 \affiliation{\Lundmed}
 \author{Jeevan Rois}
\affiliation{\LundSR}
\affiliation{\WISE}
 \author{Martin Bech}
 \affiliation{\Lundmed}
\author{Matias Kagias}
 \email{matias.kagias@fysik.lu.se}
\affiliation{\LundSR}
\affiliation{\WISE}
\affiliation{\Nanolund}

\begin{abstract}
The Talbot effect describes the emergence of periodic patterns in perturbed propagating wave fields. The effect is well studied for perturbations from structurally coherent optics such as diffraction gratings. The emergence of freeform and metaoptical designs raises the question of whether comparable behavior can also be observed from complex, non-periodic structures. Here we demonstrate that stochastic structures inspired by recent metamaterial designs, display a strong Talbot-like behavior. Re-emergence of projected wavefronts through stochastic spinodal architectures at distinct propagation distances are proven theoretically and experimentally in the visible and hard X-ray regimes. A direct application of this phenomenon is X-ray dark-field imaging for characterizing artificial and natural meso-structured materials. Our work shows that spinodal X-ray optics effectively bridge the gap between the two opposing approaches in dark-field X-ray imaging that advocate for either spatially fully coherent (i.e gratings) or incoherent (i.e diffusers) optics. This opens opportunities for exploring a new dimension in the implementation of X-ray imaging methods. Given the impact and universality of the classical Talbot effect, we expect our work to enable new opportunities for characterizing and manipulating matter.
\end{abstract}

\maketitle

\begin{figure*}[hbt!] 
\centering
{\includegraphics[width=0.9\textwidth]
{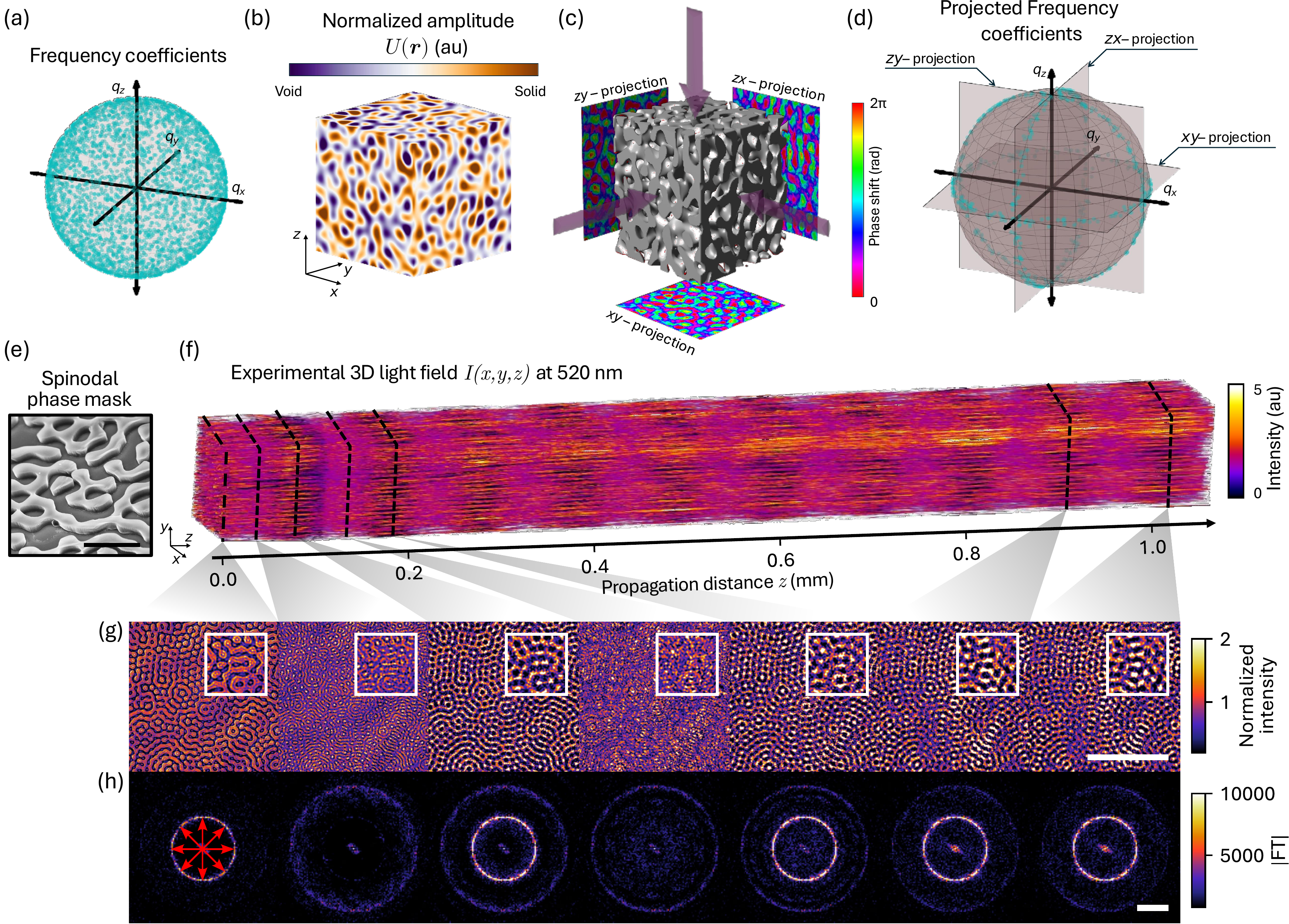}}
\caption{\textbf{Spinodal metamaterials and near field Talbot effect.} Spinodal metamaterials are defined as random Gaussian fields (b) of iso-frequency harmonic components (a)  $U(\boldsymbol{r})$. (c) Projections under any orientation yield spectrally similarly 2D spinodal optics (d). Binary PMMA spinodal optics (e) (scale bar 10 \textmu m) generate 3D light intensity patterns (f) with periodic emergence of high contrast planes. The high contrast images (g) (scale bar 100 \textmu m, zoom-in window width 50 \textmu m) correspond to self images of the spinodal optic. (h) Fourier analysis (scale bar 0.1 \textmu m$^{-1}$) further corroborates the reemergence of the spinodal patten as a ring in the frequency space.}
\label{fig:demonstration}
\end{figure*}

\begin{figure*}[hbt!] 
\centering
{\includegraphics[width=0.9\textwidth]
{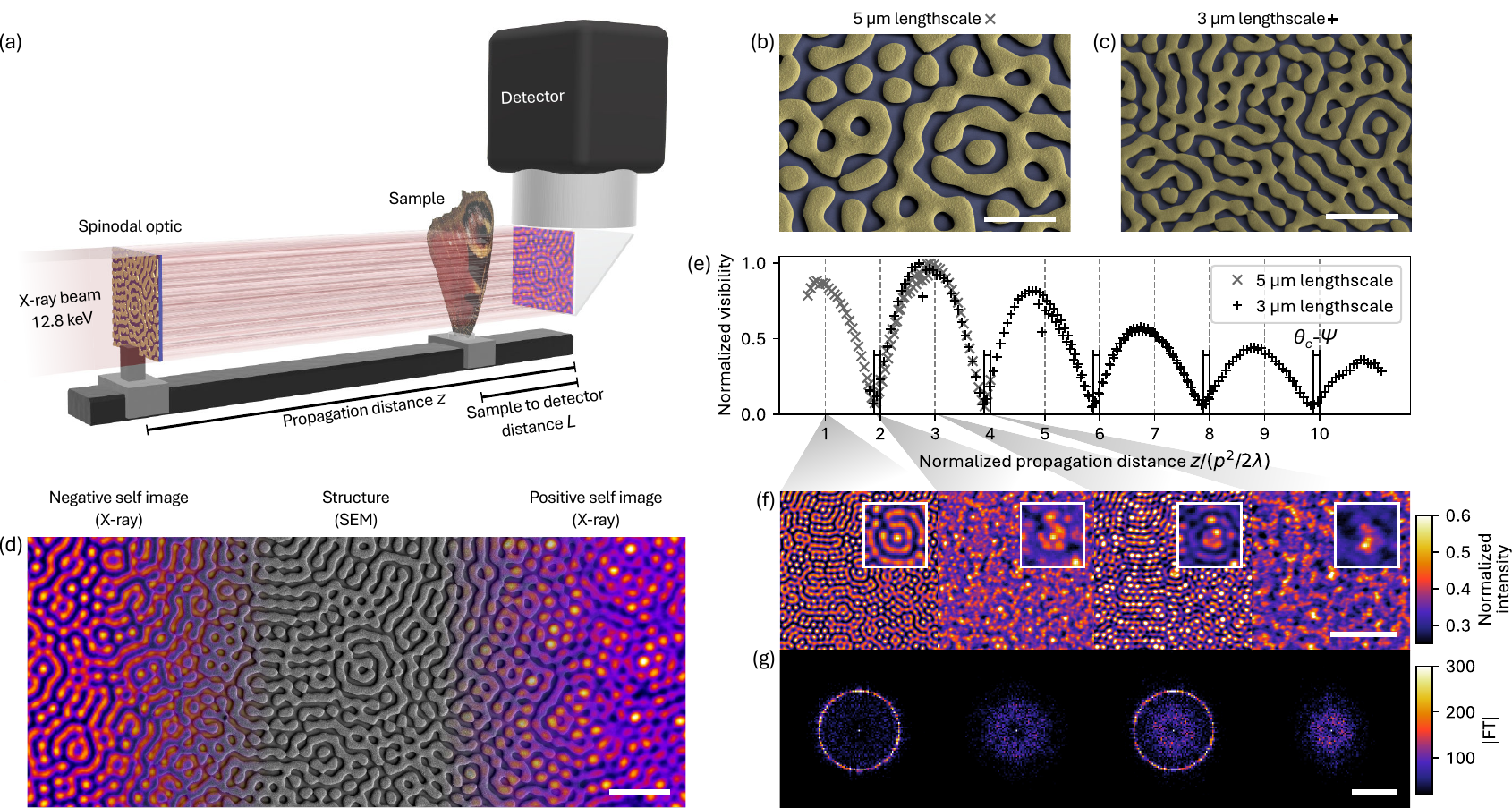}}
\caption{\textbf{Spinodal optics for X-ray dark-field imaging.} (a) Dark-field imaging setup at the ForMAX beamline comprised of linear stages and Au spinodal optics with (b) 5~\textmu m and (c) 3~\textmu m characteristic length scales to tune the autocorrelation length (scale bars 10 \textmu m). (d) Overlayed scanning electron micrograph of fabricated spinodal optical element with negative (left) and positive experimental self images (scale bar 20 \textmu m). (e) The normalized Fourier visibility shows a periodic modulation with the scaled propagation distance. The recorded intensity patterns (f) further demonstrating negative and positive self images as well as residual images (scale bar 50 \textmu m, zoom-in window width 26 \textmu m). (g) In the Fourier space, the background persists while the ring vanishes at the residual plane images (scale bar 0.2 \textmu m$^{-1}$).}
\label{fig:experimental}
\end{figure*}

When a propagating wave encounters a periodic spatial perturbation, self images of the perturbation appear at well defined repeated distances along the propagating wave vector. This is referred to as the Talbot effect \cite{talbot_lxxvi_1836, rayleigh_x_1881, berry_quantum_2001} and is universal for any system with an underlying wave formalism \cite{hu_generalized_2025} i.e electromagnetic fields, elastic waves, and wavefunctions in the form of quantum revival \cite{eberly_periodic_1980,robinett_quantum_2004}. Applications of the Talbot effect include deep ultraviolet (DUV) lithography \cite{solak_displacement_2011}, optical tweezers \cite{schonbrun_3d_2005}, quantum computing \cite{schlosser_scalable_2023}, wavefront and beam metrology at large scale facilities such as free electron lasers and synchrotron sources \cite{rutishauser_exploring_2012}, and 3D nanofabrication of metamaterials \cite{kagias_metasurface-enabled_2023}. 

A field that has significantly been accelerated by the Talbot effect in the last 20 years, is X-ray phase contrast \cite{david_differential_2002,momose_demonstration_2003,pfeiffer_phase_2006} and dark-field imaging \cite{pfeiffer_hard-x-ray_2008,yashiro_origin_2010}. Both phase-contrast and dark-field X-ray imaging have been heralded as key future technologies for bio-medical diagnostics \cite{schleede_emphysema_2012,wang_non-invasive_2014, gassert_dark-field_2025, olivo_x-ray_2025}
and material characterization \cite{revol_sub-pixel_2011,revol_laminate_2013,kim_macroscopic_2022}. Through sensitivity to coherent scattering of X-rays, these contrast modalities fundamentally offer complementary information to conventional attenuation based X-ray imaging. This has enabled high contrast X-ray imaging by mapping the phase shift of penetrated X-rays, as well as nano-microstructural contrast (dark-field) by mapping scattering from unresolved features. State-of-the-art implementations achieve this by sensing fine perturbations of spatially modulated X-ray beams. The research field has progressed in two main directions, the first advocating for fully spatially coherent (periodic) modulators (\textit{i.e} gratings) that rely on the Talbot effect, while the second exploits near field speckle patterns generated by spatially incoherent scattering media (\textit{i.e} powders or sand-paper) \cite{zanette_speckle-based_2014,berujon_near-field_2015}. Although each approach has its own benefits and limitations \cite{kashyap_experimental_2016}, one could speculate that by exploring the continuum between structural coherent modulators and incoherent modulators, instead of the extremes, new opportunities for imaging systems could emerge. In conjunction with the above mentioned unexplored space, the fields of optics and photonics have seen immense progress by departing from fully structurally coherent optics \cite{molesky_inverse_2018,lee_concurrent_2017}. The introduction of non-intuitive designs of 2D and 3D optical elements such as freeform metasurfaces has lead to new functionalities in spatiotemporal control of light \cite{shi_continuous_2020,chen_flat_2020,roberts_3d-patterned_2023}. Nonetheless, such design principles and concepts have yet to diffuse and be adopted in the field of X-ray imaging. 

Here we demonstrate both theoretically and experimentally that diffractive optics, derived from stochastic spinodal metamaterials, exhibit a Talbot-like behavior and are excellent candidates for implementing dark-field imaging systems. Spinodal metamaterials are inspired by the process of spinodal decomposition which is fundamental to several phase separation and energy minimization driven phenomena \cite{de_gennes_dynamics_1980,park_periodic_2024,kwiatkowski_da_silva_phase_2018}. Mathematically, spinodal metamaterials are synthesized as level set representations of Random Gaussian fields of harmonic components with a single frequency and amplitude but with random phase and orientation (Fig.~\ref{fig:demonstration}a--d) \cite{ma_random_2017,senhora_optimally-tailored_2022,deng_ai-enabled_2024}. Dark-field imaging experiments at the ForMAX beamline \cite{nygard_formax_2024} of the 4$^{\text{th}}$ generation synchrotron radiation facility MAX~IV demonstrate high sensitivity of the spinodal optics for retrieval of weak scattering signals from biological and artificially nanostructured materials. By bridging coherent and incoherent X‑ray modulators, spinodal optics expand the design space for flexible and robust dark‑field imaging systems and ultimately open a new direction for future developments in the field. Given the richness and universality of the Talbot effect, we expect our findings to be transferable to applications beyond the scope of this work. 

\vspace{1pc}

\noindent \textbf{Talbot effect from spinodal architectures} 
\\3D spinodal architectures are defined as level set representations of Random Gaussian fields of the following general form: $U(\boldsymbol{r}) = \sqrt{\frac{2}{N}}\sum_{n=1}^{N} \cos{(\boldsymbol{q_n}\boldsymbol{r} +\phi_{n})}$ (Fig. \ref{fig:demonstration}a). For a characteristic length scale $p = \frac{2\pi}{|\boldsymbol{q_{n}}|}$ and coherent monochromatic illumination of wavelength $\lambda$ with $\lambda \ll p$ the projection of spinodal architectures have similar spectral properties under any direction (Fig.~\ref{fig:demonstration}c--d). Assuming that the spinodal structure is comprised of a weakly absorbing material, the projected wave fields can be approximated as phase distributions of the form $e^{-i\frac{2\pi}{\lambda} \delta \tilde{P}_{\boldsymbol{s}} U(\boldsymbol{r})}$, where $\tilde{P}_{\boldsymbol{s}}$ is a projection operator in direction $\boldsymbol{s}$ and $\delta$ the refractive index decrement of the material. Given the spectral composition of $\tilde{P}_{\boldsymbol{s}}U(\boldsymbol{r})$ (single ring) we can approximate the spectrum of the exiting wave to be reasonably close to that of $\sim \tilde{P}_{\boldsymbol{s}}U(\boldsymbol{r})$ (Supplemental Note I). It is further assumed that $\tilde{P}_{\boldsymbol{s}}U(\boldsymbol{r})$ can be reasonably approximated by a binary distribution (Supplemental Note III). This means that 2D binary diffractive optics can be defined from the projection of 3D spinodal architectures.

\vspace{1pc}

\noindent The 3D wave-field intensity pattern from a poly methyl methacrylate (PMMA) spinodal phase mask (Fig.~\ref{fig:demonstration}e), shows a clear periodic modulation with propagation distance (Fig.~\ref{fig:demonstration}f and Supplemental Movie 1). High and low contrast planes are observed at repeated distances. The high contrast planes exhibit a ring in the frequency domain similar to that of the binary design of the spinodal optic. This is a strong indication of the potential Talbot-like effect. In the Supplemental Note II it is derived that strong approximate self images of $\tilde{P}_{\boldsymbol{s}}U(\boldsymbol{r})$ appear at specific distances. It is shown that the intensity pattern of a monochromatic wave field that has been modulated by passing through a spinodal binary optical element can be described as

\begin{equation}
I(x,y,z) \approx |c|^2+M(z)\tilde{P}_{\boldsymbol{s}}U(\boldsymbol{r}),
\label{eq:talbot}
\end{equation}

\noindent where $c\in \mathbb{C}$ is the average value of the complex transmission function of the spinodal element and $M(z)(M:\mathbb{R}\rightarrow \mathbb{R})$ is a function that depends on the two complex levels of the spinodal optic, the propagation distance $z$, the wavelength $\lambda$, and the characteristic length scale $p$. Specifically, $M(z)\propto \cos \left(z\frac{2 \pi \lambda}{p^2}  +\theta_{c} -\Psi  \right)$, where $\theta_{c}$ is the phase of the complex number $c$, and $\Psi$ the phase of difference of the two levels of the complex wavefront defined by the binary optical element. This means that a periodic modulation with distance is present. Wave optics simulations (Supplemental Fig. S\ref{fig:SI_simulation}a) further corroborate the appearance of strong self images. These are categorized as negative and positive depending on the sign of $M(z)$ and occur at $z_{-}^{n} = \frac{[(2n+1)\pi+\theta_{c}-\Psi]p^2}{2\pi \lambda}$ and $z_{+}^{n} =\frac{(2n\pi+\theta_{c}-\Psi)p^2}{2\pi \lambda}$, $n\in \mathbb{N}_{0}$ respectively. The periodic emergence of these self images is further demonstrated by examining the visibility (normalized spectral density) of the frequency components on the isoline $|\boldsymbol{q}|=\frac{2\pi}{p}$ (Supplemental Fig.~S\ref{fig:SI_optical}).    
\begin{figure*}[hbt!]
\centering
{\includegraphics[width=0.8\textwidth]
{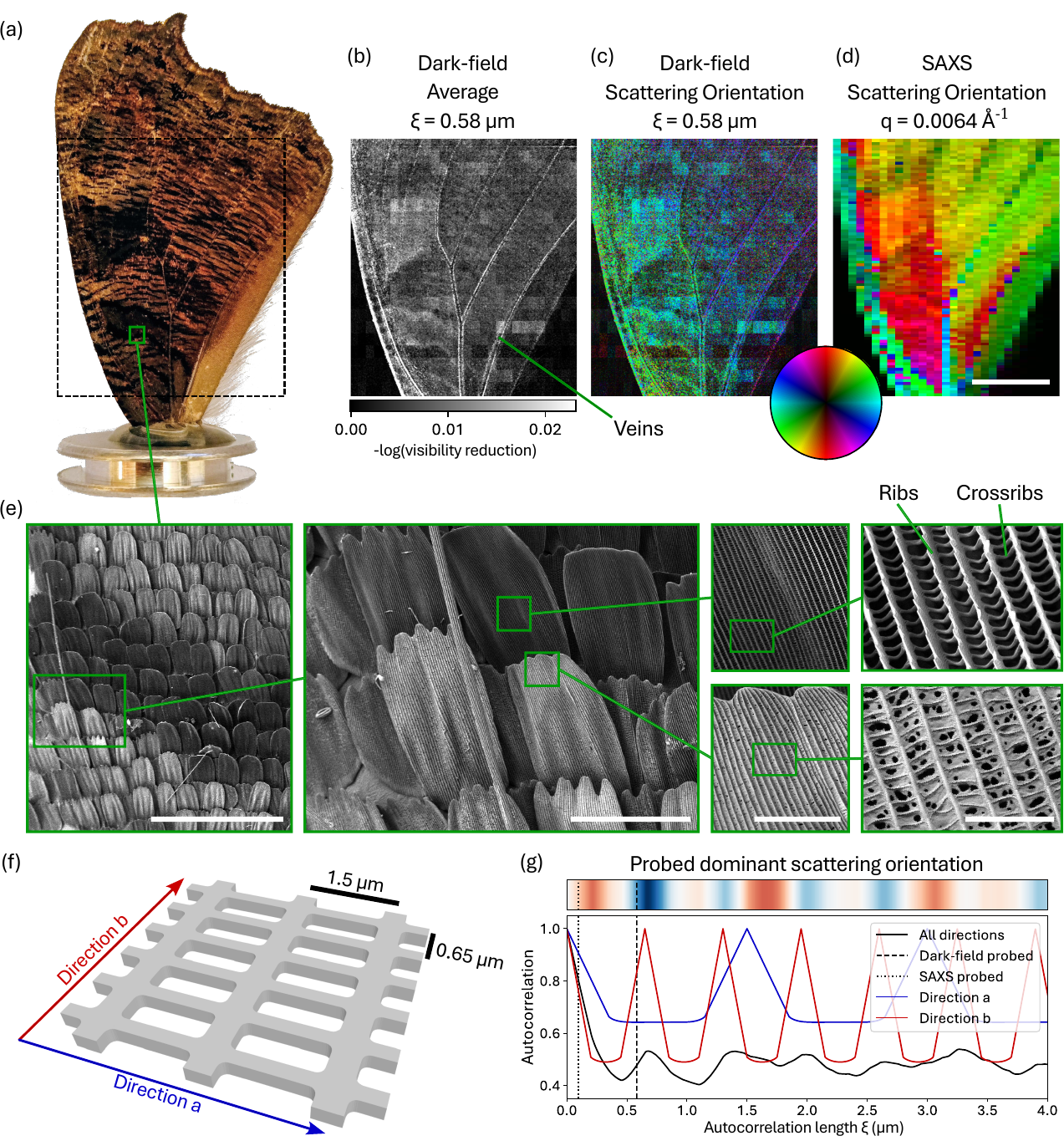}}
\caption{\textbf{Mapping morphological variations in Lepidoptera wings.} (a) Photograph of the imaged moth wing. (b)~Azimuthal averaged dark-field signal at an autocorrelation length of 580 nm. (c) Main scattering orientation of the extracted dark-field signal at the same autocorrelation length $\xi$. The color encodes the scattering direction and the intensity the azimuthal averaged dark-field signal. (d) Orientation of the SAXS data for $q=0.0064\,\text{\r{A}}^{-1}$ (scale bar 5 mm). (e) Scanning electron micrograph of a moth wing under different magnifications revealing two distinct scale morphologies. (scale bars left to right 500 \textmu m, 100 \textmu m, 20 \textmu m, 5 \textmu m). (f) A 2D simplified model of the scale anatomy can be used to interpret the delineating probed dominant scattering orientation (g).}
\label{fig:moth}
\end{figure*}

\noindent \textbf{Hard X-ray omnidirectional dark-field imaging setup with spinodal optics} 
\\A dark-field imaging setup (Fig.~\ref{fig:experimental}a) was developed for the ForMAX beamline at the 4\textsuperscript{th} generation synchrotron radiation facility MAX IV. The setup (see methods for more details) is comprised of spinodal optics, a linear stage for variable positioning of the sample, and a scintillator coupled to a sCMOS camera with a 10$\times$ optical magnification. The $\pi/2$ phase shifting spinodal X-ray optics, with a design photon energy of 12.8 keV, were fabricated by direct laser writing and gold electroplating (methods) with characteristic length scales of 3~\textmu m and 5~\textmu m, respectively (Fig. \ref{fig:experimental}b--c) reaching ranges of autocorrelation lengths $0.16~\text{\textmu m} \leq \xi_{3\text{\textmu m}}\leq 13~\text{\textmu m}$ and $0.099~\text{\textmu m}\leq\xi_{5\text{\textmu m}}\leq 9.7~\text{\textmu m}$. Both negative and positive self images were observed (Fig.~\ref{fig:experimental}d) as expected from theory, and showed excellent correspondence with the designed structures. Furthermore, periodic emergence of strong self images as well as a singular ring in the Fourier domain verified the design and optimization criterion of the imaging setup. The normalized visibility showed a decay at longer distances, which can be understood as a consequence of the limited spatial coherence of the X-ray source. The locations of visibility maxima and minima (Fig. \ref{fig:experimental}e) are slightly shifted from the nominal $np^2/2\lambda$ corresponding to $\delta\Phi=\pi/2$, $\nu=0.5$, and $\tau = 1$. This is a direct result of the fact that both the duty cycle, attenuation, and phase shift of the fabricated optics are not ideal.

Spinodal optics have several advantages compared to other modulation based approaches. First, due to the ring like spectral representation, they provide omnidirectional scattering sensitivity in the imaging plane. Secondly, they are shift and scale invariant and hence overcome the limitations of other omnidirectionally sensitive approaches utilizing arrays of optical elements such as circular gratings \cite{kagias_2d-omnidirectional_2016,kagias_diffractive_2019} and zone plates \cite{kagias_simultaneous_2021}. The scale invariance only breaks down at window sizes that approach the characteristic length scale $p$ of the fundamental function $U(\boldsymbol{r})$ (Supplemental Fig.~S\ref{fig:SI_window_test}). 

Dark-field imaging systems retrieve structural information by sampling the projected real-space autocorrelation function of the electron density of the sample \cite{strobl_general_2014} at an autocorrelation length $\xi = \frac{\lambda L }{p}$, where $L$ is the sample to detector distance. When using an X-ray imaging system based on optical magnification of a scintillator image, full-field imaging resolution is typically in the range of 1~\textmu m. Therefore, the dark-field capability of the system is ideal for characterizing structures with features in the sub-micrometer regime.

\begin{figure*}[hbt!] 
\centering
{\includegraphics[width=0.9 \textwidth]
{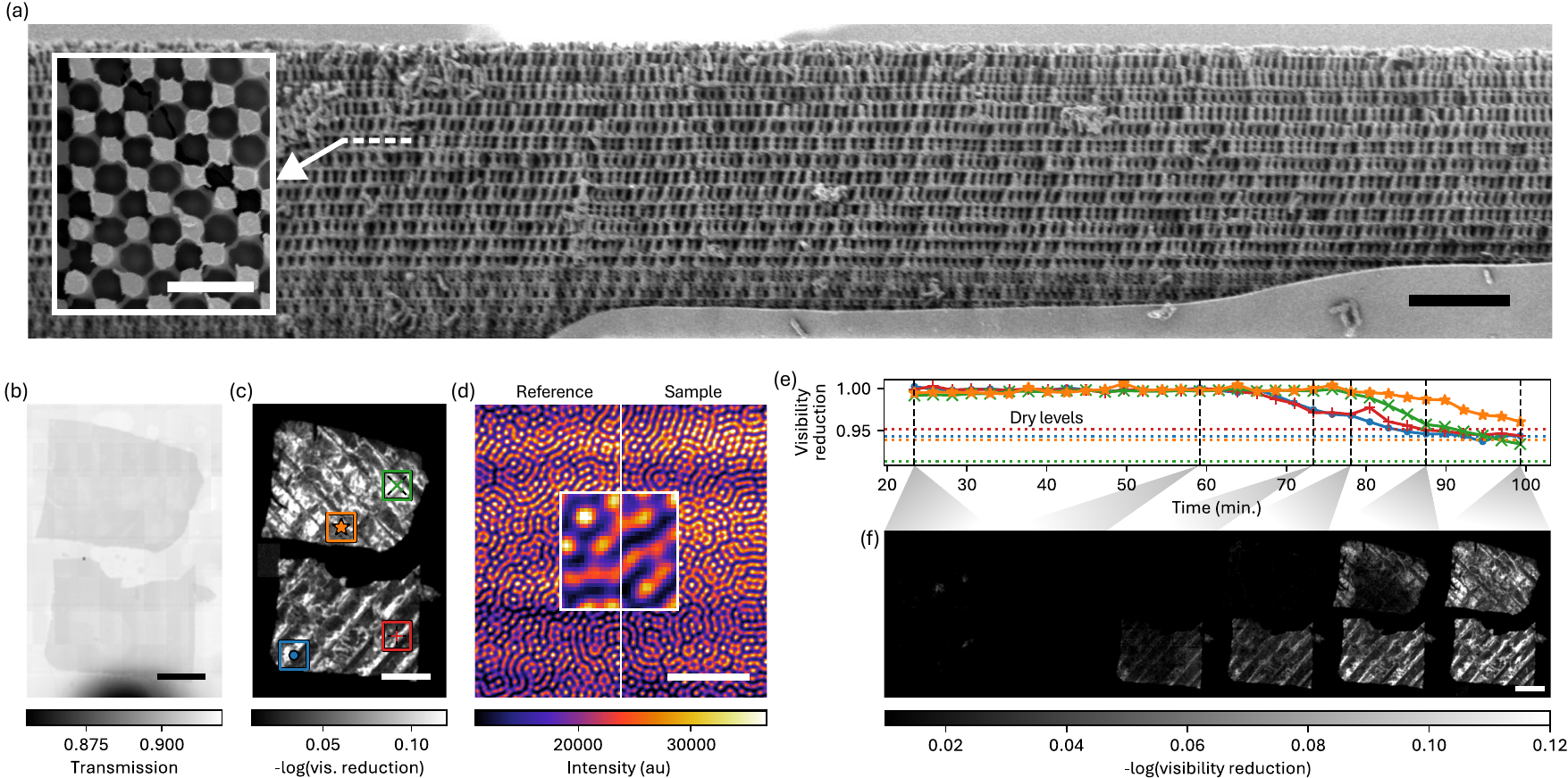}}
\caption{\textbf{Imaging of nanoarchitected metamaterials.} (a) Cross sectional scanning electron micrograph of the nanoarchitected metamaterial (scale bar 10 \textmu m, scale bar inset 2 \textmu m). (b) Averaged transmission image and (c) dry dark-field image of two nanoarchitected metamaterial sheets (scale bar 1 mm). The material is causing a slight blurring (d) in the spinodal pattern (scale bar 50 \textmu m, zoom-in window width 20 \textmu m). (e) Evolution of dark-field signal for the marked areas in (c) during a drying process within a mixture of 1:1 H$_2$O and IPA.}
\label{fig:timeseries}
\end{figure*}

\vspace{1pc}

\noindent \textbf{Probing wing anatomy in Lepidoptera.} Structural coloration of Lepidoptera wings (butterflies and moths) originates from fine morphological variations in the anatomy of the wing scales \cite{balakrishnan_nanoscale_2025}. Typical scales are composed of quasi periodic structures referred to as ribs and crossribs with thicknesses in the range of \mbox{200--400~nm}, which makes dark-field imaging tuned to sub micrometer length scales an ideal tool for large scale morphological mapping of the scale anatomy of Lepidoptera. For demonstration, we imaged the wing of a moth (Fig.~\ref{fig:moth}a) characterized by two distinct color tones. 

Both the azimuthally averaged dark-field signal at an autocorrelation length of 580~nm (Fig. \ref{fig:moth}b) and attenuation (Supplemental Fig.~S\ref{fig:SI_butterfly_transmission}) revealed the larger anatomical features of the wing such as the veins. However, the dark-field signal further reveals fine variations across the wing, originating from morphological differences at the scale level. By scanning electron microscopy (SEM), two types of scales were observed; one with open spacing between the crossribs, and one with closed spacing (Fig.~\ref{fig:moth}e). The closed scales are expected to have a lower average dark-field signal intensity, as a more modest correlation is expected. Since spinodal optics provided access to omnidirectional dark-field sensitivity, the main underlying orientation of the structure was retrieved (Fig.~\ref{fig:moth}c). From the SEM images, it is evident that the retrieved orientation is representative of the rib structure. With an approximate 2D model of the scale morphology (Fig.~\ref{fig:moth}f and Supplemental Fig.~S\ref{fig:SI_model}) we are able to estimate what main scattering orientation is most dominant at an autocorrelation length $\xi$ (Fig. \ref{fig:moth}g). A stronger scattering signal corresponds to a smaller value of the autocorrelation function. Interestingly, the orientation (color) and uncertainty (saturation) vary significantly across the range of accessible autocorrelation length of our imaging setup. In contrast, the underlying orientation retrieved from conventional scanning small angle X-ray scattering (SAXS) at a scattering vector of $q=0.0064\,\text{\r{A}}^{-1}$ (Fig.~\ref{fig:moth}d) revealed the orientation of the cross ribs which vary on a smaller length scale and correspond to a shorter autocorrelation length $\xi$. This further strengthens the notion that dark-field imaging is closing the gap between conventional full-field imaging and SAXS by providing complementary information.

\begin{figure*}[hbt!] 
\centering
{\includegraphics[width=0.8\textwidth]
{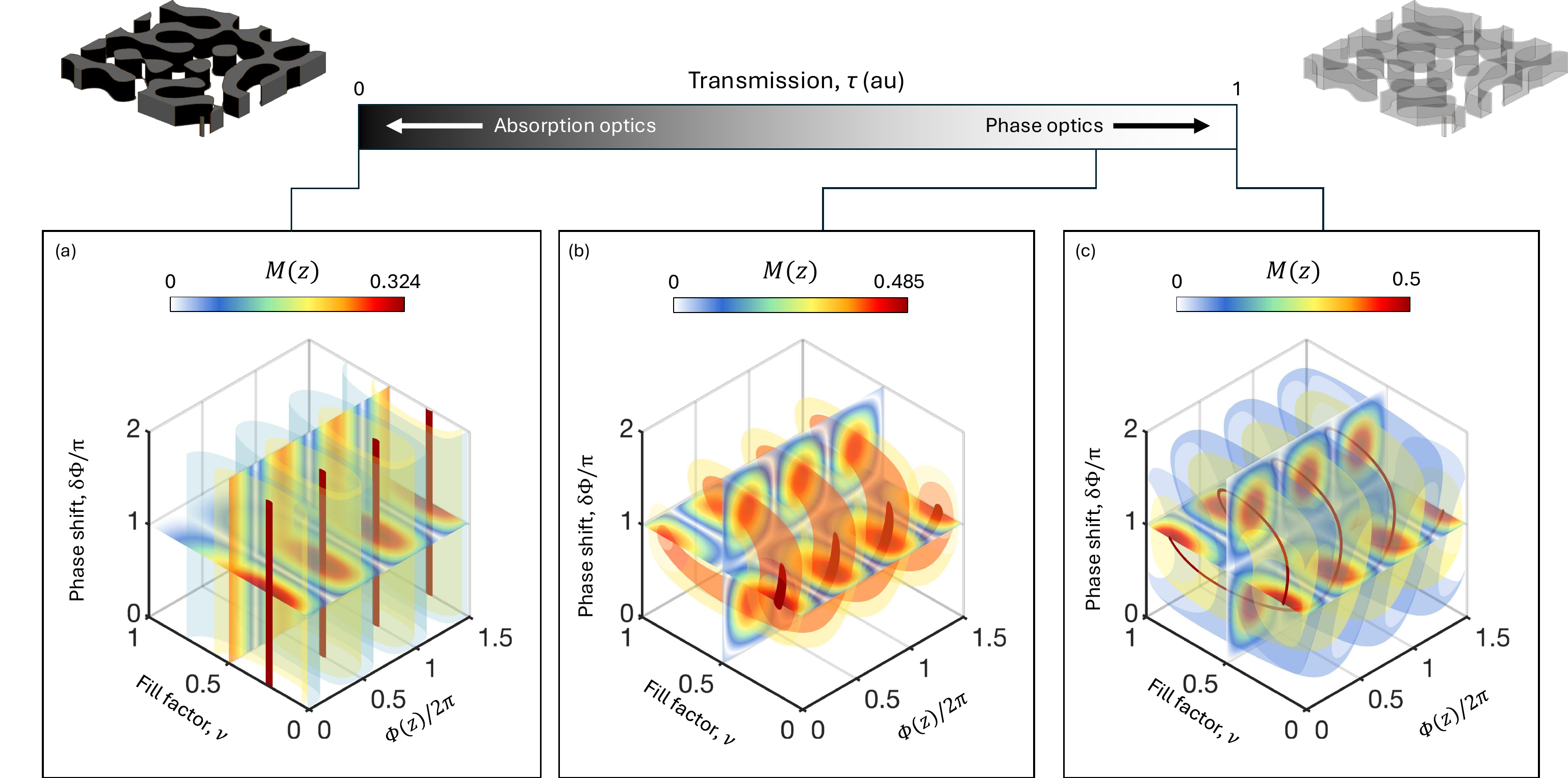}}
\caption{\textbf{{Optimization of binary spinodal optics.}} The behavior of spinodal optics can be understood as a function of the modulation term $M(z)$ in a three dimensional space comprised of the phase shift $\delta\Phi$, fill factor $\nu$, and transmission $\tau$ of the optical element. For each subplot two orthogonal slices at $\nu = 0.5$ and $\delta\Phi/\pi = 1$ as well as three isosurfaces are shown. (a) In the case of full absorbing optics modulation maxima are observed at distinct distances (iso-surfaces at 0.1, 0.2, 0.324). (b) Mixed optics ($\tau = 0.9$) also exhibit maxima at discrete distances, however sub-optimal designs with the same modulation strength can be implemented continuously along $z$ (iso-surfaces at 0.3, 0.4, 0.48). (c) In the case of pure phase shifting optics, optimal conditions can be achieved at any distance $z$ (iso-surfaces at 0.1, 0.3, 0.5).}
\label{fig:optics}
\end{figure*}

\vspace{1pc}

\noindent \textbf{Drying dynamics in 3D architected materials} Architected metamaterials, are an emerging class of structural materials with enhanced macroscopic materials properties originating from structural composition rather than chemistry \cite{xia_responsive_2022,portela_supersonic_2021}. The well defined 3D porous network of architected metamaterials render them ideal for future electrochemical, catalytic, and filtration applications. Understanding, transport properties and how they are modulated by the underlying morphology is key for further adoption of nanoarchitected metamaterials. Here, as a proof of principle, we investigated evaporation processes in scalable architected metamaterials fabricated via metasurface holographic lithography \cite{kagias_metasurface-enabled_2023}. 

The imaged structures were 25--30~\textmu m thick sheets comprised of a body center tetragonal (BCT) lattice of $\sim$500~nm diameter and 1.6 \textmu m height epoxy (SU-8) pillars (Fig.~\ref{fig:timeseries}a). Although from cross sectional SEM images, the imaged sheets appeared fairly uniform, the retrieved dark-field image at a correlation length of 290~nm, showed strong spatial variation for two pieces of architected polymer (Fig.~\ref{fig:timeseries}c). Since the probed correlation length is smaller than the pillar diameter, the average uniformity of single pillars within the analysis window (see methods) is mapped in the dark-field image. This means that our imaging approach effectively reveals slight morphological variations well within the sub-micrometer regime and maps the spatial inhomogeneity of the structure. The transmission signal shown (Fig.~\ref{fig:timeseries}b) is the average of the entire drying series (Supplemental Fig.~S\ref{fig:SI_timeseries}) and detects together with visible light images (Supplemental Fig.~S\ref{fig:SI_architected}) no apparent variation. 

The drying study was performed by wetting the nanoarchitected sheets with a 1:1 mixture of deionized water and isopropanol (IPA). The process was followed with the dark-field images taken at intervals of 141 seconds. Differences in both drying rates, and drying front propagation were observed (Fig.~\ref{fig:timeseries}e). Although we do not present a mechanistic interpretation of the probed differences in evaporation rates and sequences, we highlight that our method is highly sensitive to fluid transport in sub-micrometer pore spaces which cannot be studied with direct imaging modalities such as micro-tomography.

\vspace{1pc}

\noindent \textbf{Design space of binary spinodal optics} To further motivate the design of future dark-field imaging systems with spinodal optics, a  quantitative visibility optimization needs to be performed. Equation~\ref{eq:talbot} exhibits a significantly different behavior compared to the well established Talbot behavior from binary linear gratings (Supplemental Fig.~S\ref{fig:SI_linear_comparison}). By ignoring the higher order terms, a sinusoidal modulation described by $M(z)$ of the self images of $U(\boldsymbol{r})$ with respect to the propagation distance $z$ is derived (Supplemental Note II). The behavior of $M(z)$ in relation to the parameters of the spinodal optics (phase shift $\delta\Phi$, transmission $\tau$, fill factor $\nu$) gives an in-depth understanding of the design space of potential imaging systems with different materials and photon energies. In the most simple case of fully attenuating optics, $M(z)$ is only dependent on the fill factor $\nu$ and the propagation phase $\Phi(z) = \pi \lambda z/p^2$. Singularly repetitive maxima are observed at discrete distances (Fig.~\ref{fig:optics}a). At the other extreme, optics with only phase shifting contribution are considered. The maximum modulation is significantly higher compared to the attenuation optics. This time the visibility maximum inscribes a spiral in the $(\nu,\delta\Phi,\Phi(z))$ space (Fig.~\ref{fig:optics}c). This means that for any optics to detector distance, there is a combination of $(\nu,\delta\Phi)$ that results in the theoretical maximum visibility. This property collapses with the introduction of attenuation simultaneously with the phase shift (Fig.~\ref{fig:optics}b). However, high visibility iso-surfaces still inscribe spirals in the $(\nu,\delta\Phi,\Phi(z))$ space, hence corroborating the existence of designs with equal performance residing on a continuum.

\section*{Conclusions}\label{sec:disc}

The presence of a Talbot effect for spinodal optics inspired by metamaterials opens new possibilities for dark-field imaging modalities. Spinodal optics bridge the gap between the two prevailing camps of modulator based phase and dark-field contrast imaging. Coherent grating approaches have the benefit of well described image formation processes and clear system design and optimization guidelines. Alas, lack versatility in terms of design constrains. On the contrary, speckle based approaches, operate on a continuum and offer high flexibility in terms of system design, but lack rigorous design principles ensuring optimal imaging conditions, a feature critical in the case of biomedical and radiation dose sensitive applications. Additionally, speckle patterns are not inherently ergodic, and therefore spatial invariance is not guaranteed, leading to non uniform image retrieval performance. Spinodal optics offer a wide range of optimal design conditions with a clear optimization strategy while simultaneously adopting all the benefits of spatially stochastic modulators. Here we only demonstrate the spinodal optics in a single modulator configuration, however it is rather straightforward to envisage systems that combine multiple optics (i.e dual phase \cite{kagias_dual_2017}, Talbot-Lau \cite{pfeiffer_hard-x-ray_2008} configuration) as well as multiple length scales. This simultaneously alleviates several of the shortcomings of both modulator based imaging approaches and opens the possibility for new design strategies.

Although not explicitly demonstrated, one of the key advantages of 3D spinodal architectures is their inherent insensitivity to alignment. We have already taken concrete steps in implementing such volumetric 3D optical elements with additive manufacturing. We expect 3D spinodal optics enabled by additive manufacturing, to drive the next generation of modulator based X-ray imaging. While low photon energy ($\sim$15 keV) applications can be implemented by polymers and glassy carbon, higher energies will require high atomic number ($Z$) materials. The most promising fabrication pathway for high $Z$ volumetric optics is hydrogel ion metal additive manufacturing which has shown significant improvements both in terms of feature control, materials, and resolution reaching the sub-micrometer scale \cite{saccone_additive_2022,zhang_suppressed_2023,ji_hydrogel-based_2026}. Finally, we expect this work to inspire further progress in complex designs of 3D optics for X-ray applications beyond modulator based dark-field imaging. 

\section*{Methods}\label{sec:methods}

\noindent \textbf{X-ray optics fabrication} The spinodal phase modulators for the X-ray regime were fabricated using UV-lithography combined with electroplating. A double side polished 200~\textmu m thick [100] Si Wafer was coated with a 10~nm chromium adhesion layer and a 40~nm seed layer gold using e-beam evaporation. The substrate was solvent cleaned and dehydration baked at 115°C for 60~s. After spin-coating of the positive photoresist S1813 at 3000 rpm, the wafer was soft baked at 115°C for 60~s on a hotplate. The pattern was exposed (405 nm) using the MLA150 maskless aligner (Heidelberg Instruments) and the photoresist was developed using MF319. The structures were electroplated using the TSG-250 sulfate-based electroplating solution. The remaining photoresist was removed with organic solvents. 
The final heights for the modulators were 1.2 \textmu m and 1.3 \textmu m for the 3 \textmu m and 5 \textmu m length scale modulator, respectively (Supplemental Fig.~S\ref{fig:SI_fabrication}c).

\vspace{1pc}

\noindent \textbf{Phase mask fabrication} Spinodal optical patterns were fabricated on fused quartz substrates. Following standard solvent cleaning, the substrates were dehydrated at 150°C for 5 min and treated with oxygen plasma to improve resist adhesion. A 2.3 µm thick film of 950k PMMA A11 was spin-coated at 3000 rpm and prebaked on a hotplate at 180°C for 8 min. A conductive polymer layer (Electra 92, AR-PC 5090.02) was subsequently spin-coated at 4000 rpm and baked at 90°C for 2 min to suppress charging during electron-beam exposure. The patterns were written using a Raith Voyager electron-beam lithography system operated at 50 keV, with a write field of 500 µm, a step size of 6 nm, and an area dose of 550 µC/cm$^2$. Prior to development, the conductive layer was removed in deionized water. The PMMA was then developed in a 3:7 water/IPA solution, followed by IPA rinsing and nitrogen drying.

\vspace{1pc}

\noindent \textbf{X-ray modulator measurements} The beamline measurements were conducted at the ForMAX beamline \cite{nygard_formax_2024} at the MAX IV storage ring. A double crystal Si monochromator was used to tune the photon energy to 12.8 keV. For the full-field imaging experiments a sCMOS Andor Zyla 5.5 detector with a pixel size of 6.5~\textmu m was utilized and coupled with a 10$\times$ magnification to a 15~\textmu m thick LuAG scintillator that converted the incoming X-rays to visible light resulting in an effective pixel size of 0.65~\textmu m. The area covered by the beam on the detector was approximately 1.1~mm vertical $\times$ 0.7~mm horizontal.
Two full-field experiments were conducted. First, the spinodal modulators were mounted on a translation stage enabling to scan the propagation distance between modulators and detector, while keeping the detector in a fixed position. For both length scale of modulators this was performed with two position offsets to cover a larger range (84~mm to~518 mm) of propagation distances. The propagation distance was scanned in steps of 1~mm with an exposure time of 30 ms per image. Additionally, flat images (no sample) and dark images (no beam) were collected.
The second experimental configuration in full-field mode was used to measure a set of test samples. The 5~\textmu m length scale spinodal modulator was mounted at a distance of 140~mm to the detector. The architected material and the silk fibroin were mounted at a distance of 15~mm to the detector. This distance was increased to 30~mm for the Lepidoptera wing due to space restrictions. All samples were mounted on top of two translation stages transversal to the optical axis enabling the scanning of a larger sample area. The exposure time for each image was 15~ms. Additional to the sample images (sample and modulator) dark images were taken without X-ray beam. Each scan area was selected in a way that it contained frames without sample, later used as reference images for retrieval. For the drying series of the architected material the sample was first measured in dry state and then wetted with a 1:1 solution H$_2$O:IPA. During the following 33 images of the complete sample were acquired with a time of 2:22 min between the images (Supplemental Fig.~S\ref{fig:SI_timeseries}).

\vspace{1pc}

\noindent \textbf{SAXS measurements} For the SAXS measurements the beam was focused to a pencil-beam with a spotsize of 31~\textmu m vertical $\times$ 48~\textmu m horizontal at the sample position. The SAXS detector Eiger2 X 4M (Si) was positioned 3015~mm downstream of the sample which was mounted on the same scanning translation stages as used previously. The exposure time per point for the scanning SAXS images was 10 ms for the architected material and the silk fibroin fibers, and 100 ms for the Lepidoptera wing.

\vspace{1pc}

\noindent \textbf{Visible light measurements} A 520 nm diode laser (Thorlabs Inc.) was used to illuminate the spinodal PMMA phase optics (Supplemental Fig.~S\ref{fig:SI_setup_optical}). The generated light field after the phase mask was magnified by a Zeiss A-Plan 10$\times$ objective lens and projected on a Zelux 1.6 MP Monochrome CMOS with a pixel size of 3.45~\textmu m. 1187 images where acquired with a step size of 1~\textmu m and exposure time of 28~ms. A linear polarization filter was utilized to attenuate the 5 mW laser. Additionally, flat images and dark images were collected.

\vspace{1pc}

\noindent \textbf{Fourier visibility retrieval}
The optical and X-ray propagation measurements were processed in the same way. The acquired images $I_{\text{raw}}[z]$ for the propagation distances $z$ were flat-field corrected using flat images $F$ and dark images $D$. The result images $I[z] = (I_{\text{raw}}[z]-D)/(F-D)$ were Fourier transformed and converted to polar coordinates $\tilde{I}[z,k,\theta]$. The Fourier visibility $V_F[z]$ is extracted in the radial position $k$ corresponding to the characteristic length scale $p$ and their magnitude is azimuthal summed and normalized to the $k=0$ component.

\begin{equation}
    V_F[z] = \frac{\sum_{\theta}|\tilde{I}[z,k=\frac{2\pi}{p},\theta]|}{\sum_{\theta}|\tilde{I}[z,k=0,\theta]|}.
\end{equation}

\noindent Different offset and modulator scans were combined for the X-ray measurements requiring a normalization of the values for $V_F$. The raw data before normalization is shown in Supplemental Fig.~S\ref{fig:SI_absolute_visibility}. Both offset scans for each modulator are normalized to each other minimizing the difference of $V_F$ in the overlapping range of $z$, while setting the maximum in the covered range to 1. To combine different length scales of modulators the propagation distance $z$ is converted to a normalized propagation distance $2\lambda z/p^2$. For visualization, the datasets are subsampled in $z$ by a factor of 3 for the 3 \textmu m modulator and by a factor of 7 for the 5 \textmu m modulator.

\vspace{1pc}

\noindent \textbf{Dark-field signal retrieval} The extraction of visibility reduction $V_r$ (dark-field signal) was performed using a similar Fourier based procedure as described by Kagias et al. \cite{kagias_diffractive_2019}. A rectangular analyzing window with the size of 100 px $\times$ 100 px is scanned in steps of 10~px in two directions over the dark image corrected reference ($R$) and sample ($S$) measurement images. For each analysis window a 2D Fourier transform is calculated and transformed to polar coordinates [$k,\theta$]. The directional dark-field signal is retrieved as 

\begin{equation}
V_{r}[k,\theta] = \frac{\tilde{S}[k,\theta]}{\tilde{R}[k,\theta]}\frac{\tilde{R}[k=0,\theta]}{\tilde{S}[k=0,\theta]}.
\label{eq:retrieval}
\end{equation}

\noindent The visibility reduction is evaluated at $k=\frac{2\pi}{p}$. The azimuthal averaged dark-field signal is computed by averaging over $\theta$. The dominant scattering direction is retrieved by Fourier analysis in the $\theta$ direction, evaluating the phase of the second Fourier component (same applies to the SAXS data). The alignment of dark-field and SAXS orientation signal was confirmed by imaging a silk fibroin sample (Supplemental Fig.~S\ref{fig:SI_fibroin}). The transmission signal $T$ is calculated by $T = \tilde{S}[k=0]/ \tilde{R}[k=0]$. Due to the small beam size, several transversal sample positions were measured and stitched after retrieval based on motor positions. For the Lepidoptera scan the dark-field signal of a selection of sample positions was corrected due to instabilities in the beam by normalizing it to the signal in bordering pixels.

\vspace{1pc}

\noindent \textbf{Acknowledgments} \\
\noindent The authors acknowledge the MAX IV Laboratory for beamtime on the ForMAX beamline under proposal 20250360. Research conducted at MAX IV, a Swedish national user facility, is supported by Vetenskapsrådet (Swedish Research Council, VR) under contract 2018-07152, Vinnova (Swedish Governmental Agency for Innovation Systems) under contract 2018-04969 and Formas under contract 2019-02496. We acknowledge Lund NanoLab (LNL) for nanofabrication facilities utilized for all optics used in this work. We further acknowledge Kim Nygård, Myrto Asimakopoulou and Anuj Prajapati for experimental support during the beamtime. R.K and M.B acknowledge funding from the European Research Council (ERC) under the European Union’s Horizon 2020274 research and innovation program (Grant agreement No. 101089334). M.K acknowledges the Crafoord foundation (Grant No. 20240904) for instrumentation, the Kavli Nanoscience Institute cleanroom facilities at Caltech for the fabrication of nanoarchitected sheets, and the Swiss National Science Foundation (Grant No. P400P2 194371). This work was partially supported by the Wallenberg Initiative Materials Science for Sustainability (WISE) funded by the Knut and Alice Wallenberg Foundation. 

\vspace{1pc}

\noindent \textbf{Author contributions} \\
\noindent R.K fabricated the X-ray optics, conducted X-ray and visible light experiments, analyzed data, interpreted data, and visualized data. J.R developed the fabrication process for the visible light optics. M.B provided guidance, and contributed to the interpretation and discussion of the result. M.K conceived the study, conducted X-ray and visible light experiments, analyzed data, interpreted data, and supervised the project. The manuscript was written by R.K and M.K with input from all authors. 

\vspace{1pc}

\noindent \textbf{Competing interests} \\
\noindent The authors declare no competing interests.

\vspace{1pc}

\noindent \textbf{Data availability} \\
\noindent The data are available from the corresponding author under reasonable request.

\bibliography{collection}

@article{ji_hydrogel-based_2026,
	title = {Hydrogel-{Based} {Vat} {Photopolymerization} of {Ceramics} and {Metals} with {Low} {Shrinkages} via {Repeated} {Infusion} {Precipitation}},
	volume = {38},
	issn = {1521-4095},
	url = {https://onlinelibrary.wiley.com/doi/abs/10.1002/adma.202504951},
	doi = {10.1002/adma.202504951},
	language = {en},
	number = {8},
	urldate = {2026-03-01},
	journal = {Advanced Materials},
	author = {Ji, Yiming and Hong, Ying and Bhandari, Dhruv R. and Yee, Daryl W.},
	year = {2026},
	note = {},
	pages = {e04951},
}

@article{zhang_suppressed_2023,
	title = {Suppressed {Size} {Effect} in {Nanopillars} with {Hierarchical} {Microstructures} {Enabled} by {Nanoscale} {Additive} {Manufacturing}},
	volume = {23},
	issn = {1530-6984},
	url = {https://doi.org/10.1021/acs.nanolett.3c02309},
	doi = {10.1021/acs.nanolett.3c02309},
	number = {17},
	urldate = {2026-03-01},
	journal = {Nano Letters},
	author = {Zhang, Wenxin and Li, Zhi and Dang, Ruoqi and Tran, Thomas T. and Gallivan, Rebecca A. and Gao, Huajian and Greer, Julia R.},
	month = sep,
	year = {2023},
	note = {},
	pages = {8162--8170},
}

@article{kashyap_experimental_2016,
	title = {Experimental comparison between speckle and grating-based imaging technique using synchrotron radiation {X}-rays},
	volume = {24},
	copyright = {© 2016 Optical Society of America},
	issn = {1094-4087},
	url = {https://opg.optica.org/oe/abstract.cfm?uri=oe-24-16-18664},
	doi = {10.1364/OE.24.018664},
	language = {EN},
	number = {16},
	urldate = {2026-03-01},
	journal = {Optics Express},
	author = {Kashyap, Yogesh and Wang, Hongchang and Sawhney, Kawal},
	month = aug,
	year = {2016},
	note = {},
	pages = {18664--18673},
}

@article{yashiro_origin_2010,
	title = {On the origin of visibility contrast in x-ray {Talbot} interferometry},
	volume = {18},
	copyright = {© 2010 OSA},
	issn = {1094-4087},
	url = {https://opg.optica.org/oe/abstract.cfm?uri=oe-18-16-16890},
	doi = {10.1364/OE.18.016890},
	language = {EN},
	number = {16},
	urldate = {2026-03-01},
	journal = {Optics Express},
	author = {Yashiro, W. and Terui, Y. and Kawabata, K. and Momose, A.},
	month = aug,
	year = {2010},
	note = {},
	pages = {16890--16901},
}

@article{momose_demonstration_2003,
	title = {Demonstration of {X}-{Ray} {Talbot} {Interferometry}},
	volume = {42},
	issn = {1347-4065},
	url = {https://iopscience.iop.org/article/10.1143/JJAP.42.L866},
	doi = {10.1143/JJAP.42.L866},
	language = {en},
	number = {7B},
	urldate = {2026-03-01},
	journal = {Japanese Journal of Applied Physics},
	author = {Momose, Atsushi and Kawamoto, Shinya and Koyama, Ichiro and Hamaishi, Yoshitaka and Takai, Kengo and Suzuki, Yoshio},
	month = jul,
	year = {2003},
	note = {},
	pages = {L866},
}

@article{david_differential_2002,
	title = {Differential x-ray phase contrast imaging using a shearing interferometer},
	volume = {81},
	issn = {0003-6951},
	url = {https://doi.org/10.1063/1.1516611},
	doi = {10.1063/1.1516611},
	number = {17},
	urldate = {2026-03-01},
	journal = {Applied Physics Letters},
	author = {David, C. and Nöhammer, B. and Solak, H. H. and Ziegler, E.},
	month = oct,
	year = {2002},
	pages = {3287--3289},
}

@article{revol_laminate_2013,
	title = {Laminate fibre structure characterisation of carbon fibre-reinforced polymers by {X}-ray scatter dark field imaging with a grating interferometer},
	volume = {58},
	issn = {0963-8695},
	url = {https://www.sciencedirect.com/science/article/pii/S0963869513000686},
	doi = {10.1016/j.ndteint.2013.04.012},
	urldate = {2026-03-01},
	journal = {NDT \& E International},
	author = {Revol, Vincent and Plank, Bernhard and Kaufmann, Rolf and Kastner, Johann and Kottler, Christian and Neels, Antonia},
	month = sep,
	year = {2013},
	pages = {64--71},
}

@article{kim_macroscopic_2022,
	title = {Macroscopic mapping of microscale fibers in freeform injection molded fiber-reinforced composites using {X}-ray scattering tensor tomography},
	volume = {233},
	issn = {1359-8368},
	url = {https://www.sciencedirect.com/science/article/pii/S1359836822000221},
	doi = {10.1016/j.compositesb.2022.109634},
	urldate = {2025-05-13},
	journal = {Composites Part B: Engineering},
	author = {Kim, Jisoo and Slyamov, Azat and Lauridsen, Erik and Birkbak, Mie and Ramos, Tiago and Marone, Federica and Andreasen, Jens W. and Stampanoni, Marco and Kagias, Matias},
	month = mar,
	year = {2022},
	pages = {109634},
}

@article{saccone_additive_2022,
	title = {Additive manufacturing of micro-architected metals via hydrogel infusion},
	volume = {612},
	copyright = {2022 The Author(s), under exclusive licence to Springer Nature Limited},
	issn = {1476-4687},
	url = {https://www.nature.com/articles/s41586-022-05433-2},
	doi = {10.1038/s41586-022-05433-2},
	language = {en},
	number = {7941},
	urldate = {2025-02-04},
	journal = {Nature},
	author = {Saccone, Max A. and Gallivan, Rebecca A. and Narita, Kai and Yee, Daryl W. and Greer, Julia R.},
	month = dec,
	year = {2022},
	note = {},
	pages = {685--690},
}

@article{strobl_general_2014,
	title = {General solution for quantitative dark-field contrast imaging with grating interferometers},
	volume = {4},
	copyright = {2014 The Author(s)},
	issn = {2045-2322},
	url = {https://www.nature.com/articles/srep07243},
	doi = {10.1038/srep07243},
	language = {en},
	number = {1},
	urldate = {2026-02-25},
	journal = {Scientific Reports},
	author = {Strobl, M.},
	month = nov,
	year = {2014},
	note = {},
	pages = {7243},
}

@article{portela_supersonic_2021,
	title = {Supersonic impact resilience of nanoarchitected carbon},
	volume = {20},
	copyright = {2021 The Author(s), under exclusive licence to Springer Nature Limited},
	issn = {1476-4660},
	url = {https://www.nature.com/articles/s41563-021-01033-z},
	doi = {10.1038/s41563-021-01033-z},
	language = {en},
	number = {11},
	urldate = {2025-02-04},
	journal = {Nature Materials},
	author = {Portela, Carlos M. and Edwards, Bryce W. and Veysset, David and Sun, Yuchen and Nelson, Keith A. and Kochmann, Dennis M. and Greer, Julia R.},
	month = nov,
	year = {2021},
	note = {},
	pages = {1491--1497},
}

@article{xia_responsive_2022,
	title = {Responsive materials architected in space and time},
	volume = {7},
	copyright = {2022 Springer Nature Limited},
	issn = {2058-8437},
	url = {https://www.nature.com/articles/s41578-022-00450-z},
	doi = {10.1038/s41578-022-00450-z},
	language = {en},
	number = {9},
	urldate = {2025-12-20},
	journal = {Nature Reviews Materials},
	author = {Xia, Xiaoxing and Spadaccini, Christopher M. and Greer, Julia R.},
	month = sep,
	year = {2022},
	note = {},
	pages = {683--701},
}

@article{kagias_simultaneous_2021,
	title = {Simultaneous {Reciprocal} and {Real} {Space} {X}-{Ray} {Imaging} of {Time}-{Evolving} {Systems}},
	volume = {15},
	url = {https://link.aps.org/doi/10.1103/PhysRevApplied.15.044038},
	doi = {10.1103/PhysRevApplied.15.044038},
	number = {4},
	urldate = {2024-03-28},
	journal = {Physical Review Applied},
	author = {Kagias, Matias and Wang, Zhentian and Lovric, Goran and Jefimovs, Konstantins and Stampanoni, Marco},
	month = apr,
	year = {2021},
	note = {},
	pages = {044038},
}

@article{kagias_diffractive_2019,
	title = {Diffractive small angle {X}-ray scattering imaging for anisotropic structures},
	volume = {10},
	copyright = {2019 The Author(s)},
	issn = {2041-1723},
	url = {https://www.nature.com/articles/s41467-019-12635-2},
	doi = {10.1038/s41467-019-12635-2},
	language = {en},
	number = {1},
	urldate = {2024-01-15},
	journal = {Nature Communications},
	author = {Kagias, Matias and Wang, Zhentian and Birkbak, Mie Elholm and Lauridsen, Erik and Abis, Matteo and Lovric, Goran and Jefimovs, Konstantins and Stampanoni, Marco},
	month = nov,
	year = {2019},
	note = {},
	pages = {5130},
}

@article{kagias_2d-omnidirectional_2016,
	title = {{2D}-{Omnidirectional} {Hard}-{X}-{Ray} {Scattering} {Sensitivity} in a {Single} {Shot}},
	volume = {116},
	url = {https://link.aps.org/doi/10.1103/PhysRevLett.116.093902},
	doi = {10.1103/PhysRevLett.116.093902},
	number = {9},
	urldate = {2024-04-01},
	journal = {Physical Review Letters},
	author = {Kagias, Matias and Wang, Zhentian and Villanueva-Perez, Pablo and Jefimovs, Konstantins and Stampanoni, Marco},
	month = mar,
	year = {2016},
	note = {},
	pages = {093902},
}

@article{berujon_near-field_2015,
	title = {Near-field speckle-scanning-based x-ray imaging},
	volume = {92},
	url = {https://link.aps.org/doi/10.1103/PhysRevA.92.013837},
	doi = {10.1103/PhysRevA.92.013837},
	number = {1},
	urldate = {2026-02-25},
	journal = {Physical Review A},
	author = {Berujon, Sebastien and Ziegler, Eric},
	month = jul,
	year = {2015},
	note = {},
	pages = {013837},
}

@article{zanette_speckle-based_2014,
	title = {Speckle-{Based} {X}-{Ray} {Phase}-{Contrast} and {Dark}-{Field} {Imaging} with a {Laboratory} {Source}},
	volume = {112},
	url = {https://link.aps.org/doi/10.1103/PhysRevLett.112.253903},
	doi = {10.1103/PhysRevLett.112.253903},
	number = {25},
	urldate = {2026-02-25},
	journal = {Physical Review Letters},
	author = {Zanette, I. and Zhou, T. and Burvall, A. and Lundström, U. and Larsson, D. H. and Zdora, M. and Thibault, P. and Pfeiffer, F. and Hertz, H. M.},
	month = jun,
	year = {2014},
	note = {},
	pages = {253903},
}

@article{balakrishnan_nanoscale_2025,
	title = {Nanoscale cuticle mass density variations influenced by pigmentation in butterfly wing scales},
	volume = {16},
	copyright = {2025 The Author(s)},
	issn = {2041-1723},
	url = {https://www.nature.com/articles/s41467-025-62010-7},
	doi = {10.1038/s41467-025-62010-7},
	language = {en},
	number = {1},
	urldate = {2026-02-25},
	journal = {Nature Communications},
	author = {Balakrishnan, Deepan and Prakash, Anupama and Daurer, Benedikt J. and Finet, Cédric and Lim, Ying Chen and Shen, Zhou and Thibault, Pierre and Monteiro, Antónia and Duane Loh, N.},
	month = aug,
	year = {2025},
	note = {},
	pages = {7085},
}

@article{revol_sub-pixel_2011,
	title = {Sub-pixel porosity revealed by x-ray scatter dark field imaging},
	volume = {110},
	issn = {0021-8979},
	url = {https://doi.org/10.1063/1.3624592},
	doi = {10.1063/1.3624592},
	number = {4},
	urldate = {2026-02-25},
	journal = {Journal of Applied Physics},
	author = {Revol, V. and Jerjen, I. and Kottler, C. and Schütz, P. and Kaufmann, R. and Lüthi, T. and Sennhauser, U. and Straumann, U. and Urban, C.},
	month = aug,
	year = {2011},
	pages = {044912},
}

@article{wang_non-invasive_2014,
	title = {Non-invasive classification of microcalcifications with phase-contrast {X}-ray mammography},
	volume = {5},
	copyright = {2014 Springer Nature Limited},
	issn = {2041-1723},
	url = {https://www.nature.com/articles/ncomms4797},
	doi = {10.1038/ncomms4797},
	language = {en},
	number = {1},
	urldate = {2026-02-25},
	journal = {Nature Communications},
	author = {Wang, Zhentian and Hauser, Nik and Singer, Gad and Trippel, Mafalda and Kubik-Huch, Rahel A. and Schneider, Christof W. and Stampanoni, Marco},
	month = may,
	year = {2014},
	note = {},
	pages = {3797},
}

@article{schleede_emphysema_2012,
	title = {Emphysema diagnosis using {X}-ray dark-field imaging at a laser-driven compact synchrotron light source},
	volume = {109},
	url = {https://www.pnas.org/doi/full/10.1073/pnas.1206684109},
	doi = {10.1073/pnas.1206684109},
	number = {44},
	urldate = {2026-02-25},
	journal = {Proceedings of the National Academy of Sciences},
	author = {Schleede, Simone and Meinel, Felix G. and Bech, Martin and Herzen, Julia and Achterhold, Klaus and Potdevin, Guillaume and Malecki, Andreas and Adam-Neumair, Silvia and Thieme, Sven F. and Bamberg, Fabian and Nikolaou, Konstantin and Bohla, Alexander and Yildirim, Ali \"O. and Loewen, Roderick and Gifford, Martin and Ruth, Ronald and Eickelberg, Oliver and Reiser, Maximilian and Pfeiffer, Franz},
	month = oct,
	year = {2012},
	note = {},
	pages = {17880--17885},
}

@article{nygard_formax_2024,
	title = {{ForMAX} – a beamline for multiscale and multimodal structural characterization of hierarchical materials},
	volume = {31},
	copyright = {https://creativecommons.org/licenses/by/4.0/},
	issn = {1600-5775},
	url = {https://journals.iucr.org/s/issues/2024/02/00/rv5175/},
	doi = {10.1107/S1600577524001048},
	language = {en},
	number = {2},
	urldate = {2026-02-25},
	journal = {Journal of Synchrotron Radiation},
	author = {Nygård, K. and McDonald, S. A. and González, J. B. and Haghighat, V. and Appel, C. and Larsson, E. and Ghanbari, R. and Viljanen, M. and Silva, J. and Malki, S. and Li, Y. and Silva, V. and Weninger, C. and Engelmann, F. and Jeppsson, T. and Felcsuti, G. and Rosén, T. and Gordeyeva, K. and Söderberg, L. D. and Dierks, H. and Zhang, Y. and Yao, Z. and Yang, R. and Asimakopoulou, E. M. and Rogalinski, J. K. and Wallentin, J. and Villanueva-Perez, P. and Krüger, R. and Dreier, T. and Bech, M. and Liebi, M. and Bek, M. and Kádár, R. and Terry, A. E. and Tarawneh, H. and Ilinski, P. and Malmqvist, J. and Cerenius, Y.},
	month = mar,
	year = {2024},
	note = {},
	pages = {363--377},
}

@article{robinett_quantum_2004,
	title = {Quantum wave packet revivals},
	volume = {392},
	issn = {0370-1573},
	url = {https://www.sciencedirect.com/science/article/pii/S0370157303004381},
	doi = {10.1016/j.physrep.2003.11.002},
	number = {1},
	urldate = {2026-02-25},
	journal = {Physics Reports},
	author = {Robinett, R. W.},
	month = mar,
	year = {2004},
	pages = {1--119},
}

@article{hu_generalized_2025,
	title = {Generalized angle–orbital angular momentum {Talbot} effect and modulo mode sorting},
	volume = {19},
	copyright = {2025 The Author(s), under exclusive licence to Springer Nature Limited},
	issn = {1749-4893},
	url = {https://www.nature.com/articles/s41566-025-01622-3},
	doi = {10.1038/s41566-025-01622-3},
	language = {en},
	number = {4},
	urldate = {2026-02-25},
	journal = {Nature Photonics},
	author = {Hu, Jianqi and Eriksson, Matias and Gigan, Sylvain and Fickler, Robert},
	month = apr,
	year = {2025},
	note = {},
	pages = {392--399},
}

@article{rayleigh_x_1881,
	title = {X. {On} the electromagnetic theory of light},
	volume = {12},
	issn = {1941-5982},
	url = {https://doi.org/10.1080/14786448108627074},
	doi = {10.1080/14786448108627074},
	number = {73},
	urldate = {2026-02-25},
	journal = {The London, Edinburgh, and Dublin Philosophical Magazine and Journal of Science},
	author = {Rayleigh, Lord},
	month = aug,
	year = {1881},
	note = {},
	pages = {81--101},
}

@article{talbot_lxxvi_1836,
	title = {{LXXVI}. \textit{{Facts} relating to optical science. {No}. {IV}}},
	volume = {9},
	issn = {1941-5966, 1941-5974},
	url = {https://www.tandfonline.com/doi/full/10.1080/14786443608649032},
	doi = {10.1080/14786443608649032},
	language = {en},
	number = {56},
	urldate = {2026-02-25},
	journal = {The London, Edinburgh, and Dublin Philosophical Magazine and Journal of Science},
	author = {Talbot, H.F.},
	month = dec,
	year = {1836},
	pages = {401--407},
}

@article{berry_quantum_2001,
	title = {Quantum carpets, carpets of light},
	volume = {14},
	issn = {2058-7058},
	url = {https://doi.org/10.1088/2058-7058/14/6/30},
	doi = {10.1088/2058-7058/14/6/30},
	language = {en},
	number = {6},
	urldate = {2026-02-25},
	journal = {Physics World},
	author = {Berry, Michael and Marzoli, Irene and Schleich, Wolfgang},
	month = jun,
	year = {2001},
	pages = {39},
}

@article{eberly_periodic_1980,
	title = {Periodic {Spontaneous} {Collapse} and {Revival} in a {Simple} {Quantum} {Model}},
	volume = {44},
	url = {https://link.aps.org/doi/10.1103/PhysRevLett.44.1323},
	doi = {10.1103/PhysRevLett.44.1323},
	number = {20},
	urldate = {2026-02-25},
	journal = {Physical Review Letters},
	author = {Eberly, J. H. and Narozhny, N. B. and Sanchez-Mondragon, J. J.},
	month = may,
	year = {1980},
	note = {},
	pages = {1323--1326},
}

@article{lee_concurrent_2017,
	title = {Concurrent design of quasi-random photonic nanostructures},
	volume = {114},
	url = {https://www.pnas.org/doi/full/10.1073/pnas.1704711114},
	doi = {10.1073/pnas.1704711114},
	number = {33},
	urldate = {2026-02-24},
	journal = {Proceedings of the National Academy of Sciences},
	author = {Lee, Won-Kyu and Yu, Shuangcheng and Engel, Clifford J. and Reese, Thaddeus and Rhee, Dongjoon and Chen, Wei and Odom, Teri W.},
	month = aug,
	year = {2017},
	note = {},
	pages = {8734--8739},
}

@article{molesky_inverse_2018,
	title = {Inverse design in nanophotonics},
	volume = {12},
	copyright = {2018 Springer Nature Limited},
	issn = {1749-4893},
	url = {https://www.nature.com/articles/s41566-018-0246-9},
	doi = {10.1038/s41566-018-0246-9},
	language = {en},
	number = {11},
	urldate = {2026-02-24},
	journal = {Nature Photonics},
	author = {Molesky, Sean and Lin, Zin and Piggott, Alexander Y. and Jin, Weiliang and Vucković, Jelena and Rodriguez, Alejandro W.},
	month = nov,
	year = {2018},
	note = {},
	pages = {659--670},
}

@article{roberts_3d-patterned_2023,
	title = {{3D}-patterned inverse-designed mid-infrared metaoptics},
	volume = {14},
	copyright = {2023 The Author(s)},
	issn = {2041-1723},
	url = {https://www.nature.com/articles/s41467-023-38258-2},
	doi = {10.1038/s41467-023-38258-2},
	language = {en},
	number = {1},
	urldate = {2026-02-24},
	journal = {Nature Communications},
	author = {Roberts, Gregory and Ballew, Conner and Zheng, Tianzhe and Garcia, Juan C. and Camayd-Muñoz, Sarah and Hon, Philip W. C. and Faraon, Andrei},
	month = may,
	year = {2023},
	note = {},
	pages = {2768},
}

@article{chen_flat_2020,
	title = {Flat optics with dispersion-engineered metasurfaces},
	volume = {5},
	copyright = {2020 Springer Nature Limited},
	issn = {2058-8437},
	url = {https://www.nature.com/articles/s41578-020-0203-3},
	doi = {10.1038/s41578-020-0203-3},
	language = {en},
	number = {8},
	urldate = {2026-02-24},
	journal = {Nature Reviews Materials},
	author = {Chen, Wei Ting and Zhu, Alexander Y. and Capasso, Federico},
	month = aug,
	year = {2020},
	note = {},
	pages = {604--620},
}

@article{shi_continuous_2020,
	title = {Continuous angle-tunable birefringence with freeform metasurfaces for arbitrary polarization conversion},
	volume = {6},
	url = {https://www.science.org/doi/full/10.1126/sciadv.aba3367},
	doi = {10.1126/sciadv.aba3367},
	number = {23},
	urldate = {2026-02-24},
	journal = {Science Advances},
	author = {Shi, Zhujun and Zhu, Alexander Y. and Li, Zhaoyi and Huang, Yao-Wei and Chen, Wei Ting and Qiu, Cheng-Wei and Capasso, Federico},
	month = jun,
	year = {2020},
	note = {},
	pages = {eaba3367},
}

@article{deng_ai-enabled_2024,
	title = {{AI}-{Enabled} {Materials} {Design} of {Non}-{Periodic} {3D} {Architectures} {With} {Predictable} {Direction}-{Dependent} {Elastic} {Properties}},
	volume = {36},
	copyright = {© 2024 Wiley-VCH GmbH},
	issn = {1521-4095},
	url = {https://onlinelibrary.wiley.com/doi/abs/10.1002/adma.202308149},
	doi = {10.1002/adma.202308149},
	language = {en},
	number = {34},
	urldate = {2025-12-20},
	journal = {Advanced Materials},
	author = {Deng, Weiting and Kumar, Siddhant and Vallone, Alberto and Kochmann, Dennis M. and Greer, Julia R.},
	year = {2024},
	note = {},
	pages = {2308149},
}

@article{senhora_optimally-tailored_2022,
	title = {Optimally-{Tailored} {Spinodal} {Architected} {Materials} for {Multiscale} {Design} and {Manufacturing}},
	volume = {34},
	copyright = {© 2022 The Authors. Advanced Materials published by Wiley-VCH GmbH},
	issn = {1521-4095},
	url = {https://onlinelibrary.wiley.com/doi/abs/10.1002/adma.202109304},
	doi = {10.1002/adma.202109304},
	language = {en},
	number = {26},
	urldate = {2025-12-20},
	journal = {Advanced Materials},
	author = {Senhora, Fernando V. and Sanders, Emily D. and Paulino, Glaucio H.},
	year = {2022},
	note = {},
	pages = {2109304},
}

@article{ma_random_2017,
	title = {Random scalar fields and hyperuniformity},
	volume = {121},
	issn = {0021-8979},
	url = {https://doi.org/10.1063/1.4989492},
	doi = {10.1063/1.4989492},
	number = {24},
	urldate = {2025-12-20},
	journal = {Journal of Applied Physics},
	author = {Ma, Zheng and Torquato, Salvatore},
	month = jun,
	year = {2017},
	pages = {244904},
}

@article{de_gennes_dynamics_1980,
	title = {Dynamics of fluctuations and spinodal decomposition in polymer blends},
	volume = {72},
	issn = {0021-9606},
	url = {https://doi.org/10.1063/1.439809},
	doi = {10.1063/1.439809},
	number = {9},
	journal = {The Journal of Chemical Physics},
	author = {de Gennes, P. G.},
	month = may,
	year = {1980},
	note = {},
	pages = {4756--4763},
}

@article{kwiatkowski_da_silva_phase_2018,
	title = {Phase nucleation through confined spinodal fluctuations at crystal defects evidenced in {Fe}-{Mn} alloys},
	volume = {9},
	copyright = {2018 The Author(s)},
	issn = {2041-1723},
	url = {https://www.nature.com/articles/s41467-018-03591-4},
	doi = {10.1038/s41467-018-03591-4},
	language = {en},
	number = {1},
	urldate = {2025-12-20},
	journal = {Nature Communications},
	author = {Kwiatkowski da Silva, A. and Ponge, D. and Peng, Z. and Inden, G. and Lu, Y. and Breen, A. and Gault, B. and Raabe, D.},
	month = mar,
	year = {2018},
	note = {},
	pages = {1137},
}

@article{park_periodic_2024,
	title = {Periodic spinodal decomposition in double–strengthened medium–entropy alloy},
	volume = {15},
	copyright = {2024 The Author(s)},
	issn = {2041-1723},
	url = {https://www.nature.com/articles/s41467-024-50078-6},
	doi = {10.1038/s41467-024-50078-6},
	language = {en},
	number = {1},
	urldate = {2025-12-20},
	journal = {Nature Communications},
	author = {Park, Hyojin and Haftlang, Farahnaz and Heo, Yoon-Uk and Seol, Jae Bok and Wang, Zhijun and Kim, Hyoung Seop},
	month = jul,
	year = {2024},
	note = {},
	pages = {5757},
}

@article{pfeiffer_hard-x-ray_2008,
	title = {Hard-{X}-ray dark-field imaging using a grating interferometer},
	volume = {7},
	copyright = {2008 Springer Nature Limited},
	issn = {1476-4660},
	url = {https://www.nature.com/articles/nmat2096},
	doi = {10.1038/nmat2096},
	language = {en},
	number = {2},
	urldate = {2025-12-20},
	journal = {Nature Materials},
	author = {Pfeiffer, F. and Bech, M. and Bunk, O. and Kraft, P. and Eikenberry, E. F. and Brönnimann, Ch and Grünzweig, C. and David, C.},
	month = feb,
	year = {2008},
	note = {},
	pages = {134--137},
}

@article{pfeiffer_phase_2006,
	title = {Phase retrieval and differential phase-contrast imaging with low-brilliance {X}-ray sources},
	volume = {2},
	copyright = {2006 Springer Nature Limited},
	issn = {1745-2481},
	url = {https://www.nature.com/articles/nphys265},
	doi = {10.1038/nphys265},
	language = {en},
	number = {4},
	urldate = {2025-12-20},
	journal = {Nature Physics},
	author = {Pfeiffer, Franz and Weitkamp, Timm and Bunk, Oliver and David, Christian},
	month = apr,
	year = {2006},
	note = {},
	pages = {258--261},
}

@article{rutishauser_exploring_2012,
title = {Exploring the wavefront of hard {X}-ray free-electron laser radiation},
	volume = {3},
	copyright = {2012 Springer Nature Limited},
	issn = {2041-1723},
	url = {https://www.nature.com/articles/ncomms1950},
	doi = {10.1038/ncomms1950},
	language = {en},
	number = {1},
	urldate = {2025-12-20},
	journal = {Nature Communications},
	author = {Rutishauser, Simon and Samoylova, Liubov and Krzywinski, Jacek and Bunk, Oliver and Grünert, Jan and Sinn, Harald and Cammarata, Marco and Fritz, David M. and David, Christian},
	month = jul,
	year = {2012},
	note = {},
	pages = {947},
}

@article{kagias_metasurface-enabled_2023,
	title = {Metasurface-{Enabled} {Holographic} {Lithography} for {Impact}-{Absorbing} {Nanoarchitected} {Sheets}},
	volume = {35},
	copyright = {© 2023 Wiley-VCH GmbH},
	issn = {1521-4095},
	url = {https://onlinelibrary.wiley.com/doi/abs/10.1002/adma.202209153},
	doi = {10.1002/adma.202209153},
	language = {en},
	number = {13},
	urldate = {2024-03-27},
	journal = {Advanced Materials},
	author = {Kagias, Matias and Lee, Seola and Friedman, Andrew C. and Zheng, Tianzhe and Veysset, David and Faraon, Andrei and Greer, Julia R.},
	year = {2023},
	note = {},
	pages = {2209153},
}

@article{schonbrun_3d_2005,
	title = {{3D} interferometric optical tweezers using a single spatial light modulator},
	volume = {13},
	copyright = {© 2005 Optical Society of America},
	issn = {1094-4087},
	url = {https://opg.optica.org/oe/abstract.cfm?uri=oe-13-10-3777},
	doi = {10.1364/OPEX.13.003777},
	language = {EN},
	number = {10},
	urldate = {2025-12-20},
	journal = {Optics Express},
	author = {Schonbrun, Ethan and Piestun, Rafael and Jordan, Pamela and Cooper, Jon and Wulff, Kurt D. and Courtial, Johannes and Padgett, Miles},
	month = may,
	year = {2005},
	note = {},
	pages = {3777--3786},
}

@article{solak_displacement_2011,
	title = {Displacement {Talbot} lithography: a new method for high-resolution patterning of large areas},
	volume = {19},
	copyright = {© 2011 OSA},
	issn = {1094-4087},
	shorttitle = {Displacement {Talbot} lithography},
	url = {https://opg.optica.org/oe/abstract.cfm?uri=oe-19-11-10686},
	doi = {10.1364/OE.19.010686},
	language = {EN},
	number = {11},
	urldate = {2025-12-20},
	journal = {Optics Express},
	author = {Solak, Harun H. and Dais, Christian and Clube, Francis},
	month = may,
	year = {2011},
	note = {},
	pages = {10686--10691},
}

@article{schlosser_scalable_2023,
	title = {Scalable {Multilayer} {Architecture} of {Assembled} {Single}-{Atom} {Qubit} {Arrays} in a {Three}-{Dimensional} {Talbot} {Tweezer} {Lattice}},
	volume = {130},
	url = {https://link.aps.org/doi/10.1103/PhysRevLett.130.180601},
	doi = {10.1103/PhysRevLett.130.180601},
	number = {18},
	urldate = {2025-12-20},
	journal = {Physical Review Letters},
	author = {Schlosser, Malte and Tichelmann, Sascha and Schäffner, Dominik and de Mello, Daniel Ohl and Hambach, Moritz and Schütz, Jan and Birkl, Gerhard},
	month = may,
	year = {2023},
	note = {},
	pages = {180601},
}

@article{gassert_dark-field_2025,
	title = {Dark-Field Chest Radiography for Pneumothorax Detection: A                     Prospective Study},
	volume = {7},
	url = {https://pubs.rsna.org/doi/abs/10.1148/ryct.240560},
	doi = {10.1148/ryct.240560},
	shorttitle = {Dark-Field Chest Radiography for Pneumothorax Detection},
	pages = {e240560},
	number = {6},
	journaltitle = {Radiology: Cardiothoracic Imaging},
	author = {Gassert, Florian T. and Bast, Henriette and Urban, Theresa and Schick, Rafael and Lochschmidt, Maximilian E. and Kaster, Lennard and Koehler, Thomas and Karrer, Alexandra and Keppler, Ariane and Steinhardt, Manuel and Marka, Alexander W. and Steinhelfer, Lisa and Sauter, Andreas P. and Makowski, Marcus R. and Pfeiffer, Franz and Pfeiffer, Daniela},
	urldate = {2026-03-03},
	date = {2025-12},
	note = {},
}

@article{olivo_x-ray_2025,
	title = {X-ray phase contrast for intra-operative specimen imaging in breast conserving surgery and other areas},
	volume = {3},
	issn = {2813-687X},
	url = {https://www.frontiersin.org/journals/medical-engineering/articles/10.3389/fmede.2025.1608247/full},
	doi = {10.3389/fmede.2025.1608247},
	journaltitle = {Frontiers in Medical Engineering},
	shortjournal = {Front. Med. Eng.},
	author = {Olivo, Alessandro and Leff, Daniel R.},
	urldate = {2026-03-03},
	date = {2025-09-17},
	note = {},
}

@article{kagias_dual_2017,
    title = {Dual phase grating interferometer for tunable dark-field           sensitivity},
    volume = {110},
    issn = {0003-6951},
    url = {https://aip.scitation.org/doi/full/10.1063/1.4973520},
    doi = {10.1063/1.4973520},
    number = {1},
    urldate = {2023-03-23},
    journal = {Applied Physics Letters},
    author = {Kagias, Matias and Wang, Zhentian and Jefimovs, Konstantins and Stampanoni, Marco},
    month = jan,
    year = {2017},
    note = {Publisher: American Institute of Physics},
    pages = {014105},
}

\clearpage

\setcounter{equation}{0}
\setcounter{figure}{0}
\renewcommand{\thefigure}{\arabic{figure}}
\makeatletter
\renewcommand{\fnum@figure}{FIG.~S\thefigure}
\makeatother

\onecolumngrid
\begin{center}
{\Large Supplementary Information for\\[1ex] \textbf{Breaking order: Talbot effect with spinodal architectures}}\\[1ex]

{ Robin Kr\"uger$^{1}$, Jeevan Rois $^{2,3}$, Martin Bech $^{1}$, Matias Kagias$^{2,3,4}$\\ }
{\small \em $^1$ \Lundmed\\
$^2$ \LundSR\\
$^3$ \WISE\\
$^4$ \Nanolund}
\end{center}

\section{Spectral approximation for spinodal optics}

\noindent A spinodal 2D continuous function is defined as the following sum

\begin{equation}
    U(\boldsymbol{r}) =\sqrt{\frac{2}{N}} \sum_{n=1}^{N}{ \cos(\boldsymbol{q_{n}}\boldsymbol{r}+\phi_{n})}
\end{equation}

\noindent where $N\gg1$. From this we define an ideal continuous spinodal 2D optical element with a phase shift $\delta \Phi$ as

\begin{equation}
    G(\boldsymbol{r}) = e^{i\frac{\delta\Phi}{2} U(\boldsymbol{r})}
\end{equation}

\noindent This can be rewritten as

\begin{equation}
    G(\boldsymbol{r}) =  \prod_{n=1}^{N}e^{i\frac{\delta\Phi}{2} \sqrt{\frac{2}{N}} \cos(\boldsymbol{q_{n}}\boldsymbol{r}+\phi_{n})}
\end{equation}

\noindent By making use of the Jacobi-Anger expansion which states that

\begin{equation}
   e^{i\Psi \cos{u}} = \sum_{m=-\infty}^{+\infty} (i)^{m} J_{m}(\Psi)e^{imu}
\end{equation}

\noindent the spinodal optic transmission function is rewritten as
\begin{equation}
    G(\boldsymbol{r}) =  \prod_{n=1}^{N}\sum_{m=-\infty}^{+\infty} (i)^{m} J_{m}\left( \frac{\delta\Phi}{2}\sqrt{\frac{2}{N}} \right ) e^{im(\boldsymbol{q_{n}}\boldsymbol{r}+\phi_{n})}
\end{equation}

\noindent When $N\gg1$ then $\frac{\delta\Phi}{2}\sqrt{\frac{2}{N}}\ll1$ so terms with $|m|>1$ can be ignored. Making use of $J_{-1}(x)=-J_{1}(x)$ we get the following expression

\begin{equation}
    G(\boldsymbol{r}) =  \prod_{n=1}^{N} \left [ 2iJ_{1}\left( \frac{\delta\Phi}{2}\sqrt{\frac{2}{N}} \right ) \cos{(\boldsymbol{q_{n}}\boldsymbol{r}+\phi_{n})}  +J_{0} \left(\frac{\delta\Phi}{2}\sqrt{\frac{2}{N}} \right ) \right]
    \label{eq:bessel_aprox}
\end{equation}

\begin{figure*}[t] 
\centering
{\includegraphics[width=1\textwidth]{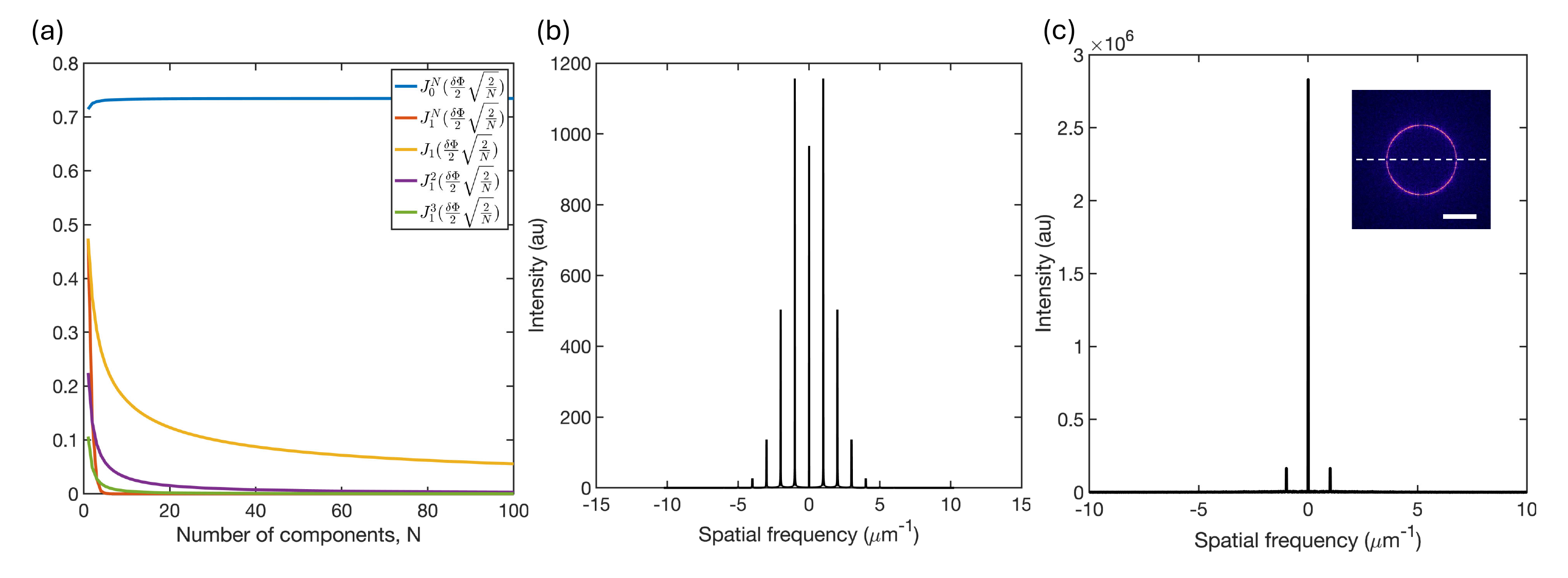}}
\caption{ \textbf{Spectral properties of spinodal phase optics.} (a) For increasing number of components $N$ of the spinodal fundamental function $U(\boldsymbol{r})$, the zeros frequency term in equation~\ref{eq:bessel_aprox} saturates (blue curve). While higher powers of the fist Bessel function which are present in $K(\boldsymbol{r})$ decay fast (red, purple, and green curves). The modulation of the spinodal function also decays but at a much slower rate (orange curve). (b) A continuous phase shifting grating $e^{i\delta\Phi/2 \cos(2\pi x/p)}$ shows multiple peaks in the frequency domain, in contrast to spinodal continuous optics that have a singular ring defined by the characteristic length scale. (c) Line profile of 2D spectrum (inset scale bar $1~\text{\textmu m}^{-1}$) of spinodal phase optic.}
\label{fig:SI_bessel}
\end{figure*}

\noindent After algebraic expansion, the following expression is reached 

\begin{equation}
    G(\boldsymbol{r}) = J_{0}^{N} \left( \frac{\delta\Phi}{2}\sqrt{\frac{2}{N}} \right )+2iJ_{0}^{N-1} \left( \frac{\delta\Phi}{2}\sqrt{\frac{2}{N}} \right )J_{1} \left( \frac{\delta\Phi}{2}\sqrt{\frac{2}{N}} \right )
    \sum_{n=1}^{N} \cos{(\boldsymbol{q_{n}}\boldsymbol{r}+\phi_{n})} + K(\boldsymbol{r})
\end{equation}

\noindent where the $K(\boldsymbol{r})$ term is given by 

\begin{equation}
    K(\boldsymbol{r}) = \sum_{l=2}^{N} \left [2iJ_{1} \left( \frac{\delta\Phi}{2}\sqrt{\frac{2}{N}} \right )  \right ]^{l} J_{0}^{N-l} \left( \frac{\delta\Phi}{2}\sqrt{\frac{2}{N}} \right ) \sum _{j=2}^{\binom Nl}\prod_{t \in S_{l}} \cos{(\boldsymbol{q_{t}}\boldsymbol{r}+\phi_{t})}
\end{equation}

\noindent where $S_{l}$ is a subset containing $l$ unique natural numbers sampled from the interval $[2,N]$. Since $K(\boldsymbol{r})$ is a sum of powers of the Bessel terms of the first kind, it will decay for high values of $N$ (Fig. S\ref{fig:SI_bessel}a) and therefore for the continuous case, the spectrum of spinodal phase optics can be described by a single ring. This is in contrast to conventional continuous linear phase gratings of the form $e^{i\delta\Phi/2 \cos(2\pi x/p)}$ that generate higher orders (Fig. S\ref{fig:SI_bessel}b and S\ref{fig:SI_bessel}c).

\section{Theoretical modeling of near field interference from Spinodal architectures} 

\noindent Let $\hat{G}(\boldsymbol{q})$ be the Fourier transform of the real space transmission function $G(\boldsymbol{r})$ ($G:\mathbb{R}^2\rightarrow \mathbb{C}$) of the spinodal optical element. The spinodal element is defined by first creating a random fundamental function $U(\boldsymbol{r})$ ($U:\mathbb{R}^2\rightarrow \mathbb{R}$) given by

\begin{equation}
    U(\boldsymbol{r}) = \sqrt{\frac{2}{N}}\sum_{n=1}^{N}{ \cos(\boldsymbol{q_{n}}\boldsymbol{r}+\phi_{n})}
\end{equation}

\noindent Where $\boldsymbol{q}_{n}$ are vectors with $|\boldsymbol{q}_{n}| = 2\pi\frac{1}{p}$ distributed on a circle with angular spacing $\frac{2\pi}{N-1}$, and $\phi_{n}$ are random phase values with $\phi_{n} \in [0,2\pi)$. From this fundamental function we generate a binary optical element that without loss of generality it can be described by using the following rule

\[
    G(\boldsymbol{r})= 
\begin{dcases}
    \tau e^{i\delta\Phi},& \text{if } U(\boldsymbol{r})\geq t\\
    1,              & \text{otherwise}
\end{dcases}
\]
\noindent where $t \in \mathbb{R}$ represents a threshold value that can be varied to create patterns with different duty cycles, $\tau$ is the transmission induced by the binary structure, $\delta \Phi$ the phase shift. For $N\gg1$ it can be assumed that the Fourier transforms of the fundamental function $\hat{U}(\boldsymbol{q})$ and the binary optical element $\hat{G}(\boldsymbol{q})$ (a phase shifting version of this) are closely related as follows:

\begin{equation}
    \hat{G}(\boldsymbol{q}) \approx c\delta(\boldsymbol{q})+w(q)\hat{U}(\boldsymbol{q})
\end{equation}

\noindent where $w\in\mathbb{C}$ is the modulation at each frequency with $|w(q)| = |w|$, $w(q) = w(-q)^{*}e^{i2\Psi}$, and $\Psi = \arctan \left (  \frac{\tau\sin(\delta\Phi)}{\tau\cos(\delta \Phi)-1} \right )$. Hence, the Fourier transform of the binary spinodal optical element is written as

\begin{equation}
    \hat{G}(\boldsymbol{q}) = c\delta(\boldsymbol{q})+|w|\sum_{n=1}^{N} \left [{\delta(\boldsymbol{q}-\boldsymbol{q_{n}})e^{-i\phi_{n}}}e^{i\Psi}+\delta(\boldsymbol{q}+\boldsymbol{q_{n}})e^{i\phi_{n}}e^{i\Psi} \right ]
    \label{eq:grat}
\end{equation}

\noindent where $c\in \mathbb{C}$ is the average value of the complex transmission function 

\begin{equation}
    c = \frac{1}{\Omega}\int_{\Omega} G(\boldsymbol{r}) \text{d}\boldsymbol{r} = |c|e^{i\theta_{c}} = (1-\nu)+ \nu\tau e^{i\delta \Phi}
\end{equation}

\noindent with 

\begin{equation}
    \theta_{c} = \arctan{\left ( \frac{-\nu\tau \sin{\delta \Phi}}{1-\nu+\nu\tau\cos{\delta \Phi}} \right )},\quad |c| = \sqrt{(1-\nu)^2+\nu^2\tau^2+2\nu(1-\nu)\tau\cos(\delta\Phi)}\label{eq:c}
\end{equation}

\noindent and $\nu \in \mathbb{R}$ is the duty cycle of the binary spinodal structure and defined as the following 

\begin{equation}
    \nu = \frac{1}{\Omega}\int_{\Omega} \frac{G(\boldsymbol{r})-1}{\tau e^{i\delta \Phi}-1}\text{~d}\boldsymbol{r} = \frac{c-1}{\tau e^{i\delta \Phi}-1}
\end{equation}

\noindent Using Parseval's theorem, we can estimate $|w|$ as 

\begin{equation}
    |w| = \sqrt{ \frac{1}{N} \left [  \nu\tau^2+(1-\nu)  - |c|^2) \right ]}
    \label{eq:w}
\end{equation}

\noindent Assuming plane wave illumination at the interaction plane with the optical element $E(\boldsymbol{r}) = 1$ the output wave is given by

\begin{equation}
    E(\boldsymbol{r},0) = G(\boldsymbol{r})E(\boldsymbol{r})
\end{equation}

\noindent In the case of hard X-rays, we use the small angle approximation that allows us to describe the wave propagation through the Fourier representation of the Fresnel propagator $\hat{H}(\boldsymbol{q},z)$ given by

\begin{equation}
    \hat{H}(\boldsymbol{q},z) = e^{-i z\frac{|\boldsymbol{q}|^2}{2k}}
\end{equation}

\noindent where $k$ is the wavevector $\frac{2\pi}{\lambda}$. The propagated field at a distance $z$ is given by

\begin{equation}
    E(\boldsymbol{r},z) = \int \hat{G}(\boldsymbol{q}) \hat{H}(\boldsymbol{q},z) e^{-i\boldsymbol{q}\boldsymbol{r}} \text{d}\boldsymbol{q}
    \label{eq:field_propagator}
\end{equation}

\noindent By substituting equation \ref{eq:grat} into \ref{eq:field_propagator} and making use of the well known property $f(x_{0}) = \int \delta(x-x_{0})f(x)\text{d}x$ as well as $\cos{x} = \frac{e^{-ix}+e^{ix}}{2}$ we end up with

\begin{equation}
    E(\boldsymbol{r},z) = c+|w|e^{-i\Phi(z)}\sum_{n=1}^{N} e^{-i(\boldsymbol{q_{n}}\boldsymbol{r}+\phi_{n}-\Psi)}+e^{i(\boldsymbol{q_{n}}\boldsymbol{r}+\phi_{n}+\Psi)}  
\end{equation}

\noindent Where $\Phi(z)$ is the phase of the Fresnel propagator given by $\pi\lambda \frac{1}{p^2} z$. Please note that $\Phi(z)$ is independent of the index $n$, since we have used the fact that $|\boldsymbol{q}_{n}| = 2\pi\frac{1}{p}$. The detected intensity $I(\boldsymbol{r},z) = E(\boldsymbol{r},z)E^*(\boldsymbol{r},z)$ of the wave field can then be written as

\begin{align}
    I(\boldsymbol{r},z) &= \left(c+|w|e^{-i\Phi(z)}\sum_{n=1}^{N} \left[e^{-i(\boldsymbol{q_{n}}\boldsymbol{r}+\phi_{n}-\Psi)}+e^{i(\boldsymbol{q_{n}}\boldsymbol{r}+\phi_{n}+\Psi)}\right]\right) \notag\\ 
    &\quad\quad\quad\times \left(c^*+|w|e^{i\Phi(z)}\sum_{n=1}^{N} \left[e^{i(\boldsymbol{q_{n}}\boldsymbol{r}+\phi_{n}-\Psi)}+e^{-i(\boldsymbol{q_{n}}\boldsymbol{r}+\phi_{n}+\Psi)}\right]\right)\notag\\
    &= |c|^{2}+|c||w|e^{i(\Phi(z)+\theta_c)}\sum_{n=1}^{N} \left[e^{i(\boldsymbol{q_{n}}\boldsymbol{r}+\phi_{n}-\Psi)}+e^{-i(\boldsymbol{q_{n}}\boldsymbol{r}+\phi_{n}+\Psi)}\right]\notag\\
    &\quad\quad\quad+|c||w|e^{-i(\Phi(z)+\theta_c)}\sum_{n=1}^{N} \left[e^{-i(\boldsymbol{q_{n}}\boldsymbol{r}+\phi_{n}-\Psi)} +e^{i(\boldsymbol{q_{n}}\boldsymbol{r}+\phi_{n}+\Psi)}\right] + O(\boldsymbol{r})\notag\\
 &=|c|^{2}+|w||c|\sum_{n=1}^{N}\left[e^{i(\boldsymbol{q_{n}}\boldsymbol{r}+\phi_{n}-\Psi+\Phi(z)+\theta_c)}+e^{-i(\boldsymbol{q_{n}}\boldsymbol{r}+\phi_{n}-\Psi+\Phi(z)+\theta_c)}\right.\notag\\&\quad\quad\quad + \left. e^{-i(\boldsymbol{q_{n}}\boldsymbol{r}+\phi_{n}+\Psi-\Phi(z)-\theta_c)}+e^{i(\boldsymbol{q_{n}}\boldsymbol{r}+\phi_{n}+\Psi-\Phi(z)-\theta_c)} \right] + O(\boldsymbol{r})\notag\\
 &=|c|^{2}+2|w||c|\sum_{n=1}^{N}\left[\cos(\boldsymbol{q_{n}}\boldsymbol{r}+\phi_n-\Psi+\Phi(z)+\theta_c) + \cos(\boldsymbol{q_{n}}\boldsymbol{r}+\phi_n+\Psi-\Phi(z)-\theta_c) \right]+ O(\boldsymbol{r})\notag\\
&=|c|^{2}+2|w||c|\sum_{n=1}^{N}\left[2\cos\left(\frac{\boldsymbol{q_{n}}\boldsymbol{r}+\phi_n-\Psi+\Phi(z)+\theta_c+\boldsymbol{q_{n}}\boldsymbol{r}+\phi_n+\Psi-\Phi(z)-\theta_c}{2}\right)\right.\notag\\
&\quad\quad\quad\times \left.\cos\left(\frac{\boldsymbol{q_{n}}\boldsymbol{r}+\phi_n-\Psi+\Phi(z)+\theta_c-\boldsymbol{q_{n}}\boldsymbol{r}-\phi_n-\Psi+\Phi(z)+\theta_c}{2}\right) \right]+ O(\boldsymbol{r})\notag\\
&= |c|^{2}+4|w||c|\cos(\Phi(z)-\Psi+\theta_c) \sum_{n=1}^{N} \cos(\boldsymbol{q_{n}}\boldsymbol{r}+\phi_n)+ O(\boldsymbol{r})\notag\\
&= |c|^{2}+|w|T(z)\sum_{n=1}^{N} \cos(\boldsymbol{q_{n}}\boldsymbol{r}+\phi_{n}) + O(\boldsymbol{r})
\label{eq:intensity}
\end{align}

\noindent Where $O(\boldsymbol{r})$ is a higher order term given by

\begin{equation}
    O(\boldsymbol{r}) = |w|^{2} \left ( \sum_{n=1}^{N} e^{-i(\boldsymbol{q_{n}}\boldsymbol{r}+\phi_{n}-\Psi)}+e^{i(\boldsymbol{q_{n}}\boldsymbol{r}+\phi_{n}+\Psi)}  \right ) \left (\sum_{n=1}^{N} e^{i(\boldsymbol{q_{n}}\boldsymbol{r}+\phi_{n}-\Psi)}+e^{-i(\boldsymbol{q_{n}}\boldsymbol{r}+\phi_{n}+\Psi)} \right )
\end{equation}

\noindent and $T(z)$ is 

\begin{align}
    T(z) = 4|c|\cos{(\Phi(z)-\Psi+\theta_{c})}\label{eq:modulation}
\end{align}

\noindent Since $N\gg1$ it follows from equation \ref{eq:w} that $|w|^2\ll|c|^2$ and therefore for larger values of $T(z)$ the higher order term $O(\boldsymbol{r})$ can be neglected meaning that self images are observed. Given this assumption, we can rewrite the intensity distribution as

\begin{equation}
 I(\boldsymbol{r},z) \approx |c|^2+4|c||w|\cos{(\Phi(z)-\Psi+\theta_{c})} \sum_{n=1}^{N} \cos(\boldsymbol{r}\boldsymbol{q_{n}}+\phi_{n})
\end{equation}

\noindent Furthermore, the above expression gives an analytical relationship to maximize the contrast of the observed self images for any duty cycle $\nu$, phase shift $\delta\Phi$ and transmission $\tau$ of the spinodal optical element. The appearance of self images at different distances can be described with the modulation function $M(z)$

\begin{equation}
    M(z) = |c||w|\cos \left(z\frac{2 \pi \lambda}{p^2}  +\theta_{c} -\Psi  \right)
    \label{eq:Mz}
\end{equation}

\section{Numerical investigation of spectral approximation of spinodal binary optics}

\begin{figure*}[h!] 
\centering
{\includegraphics[width=1\textwidth]{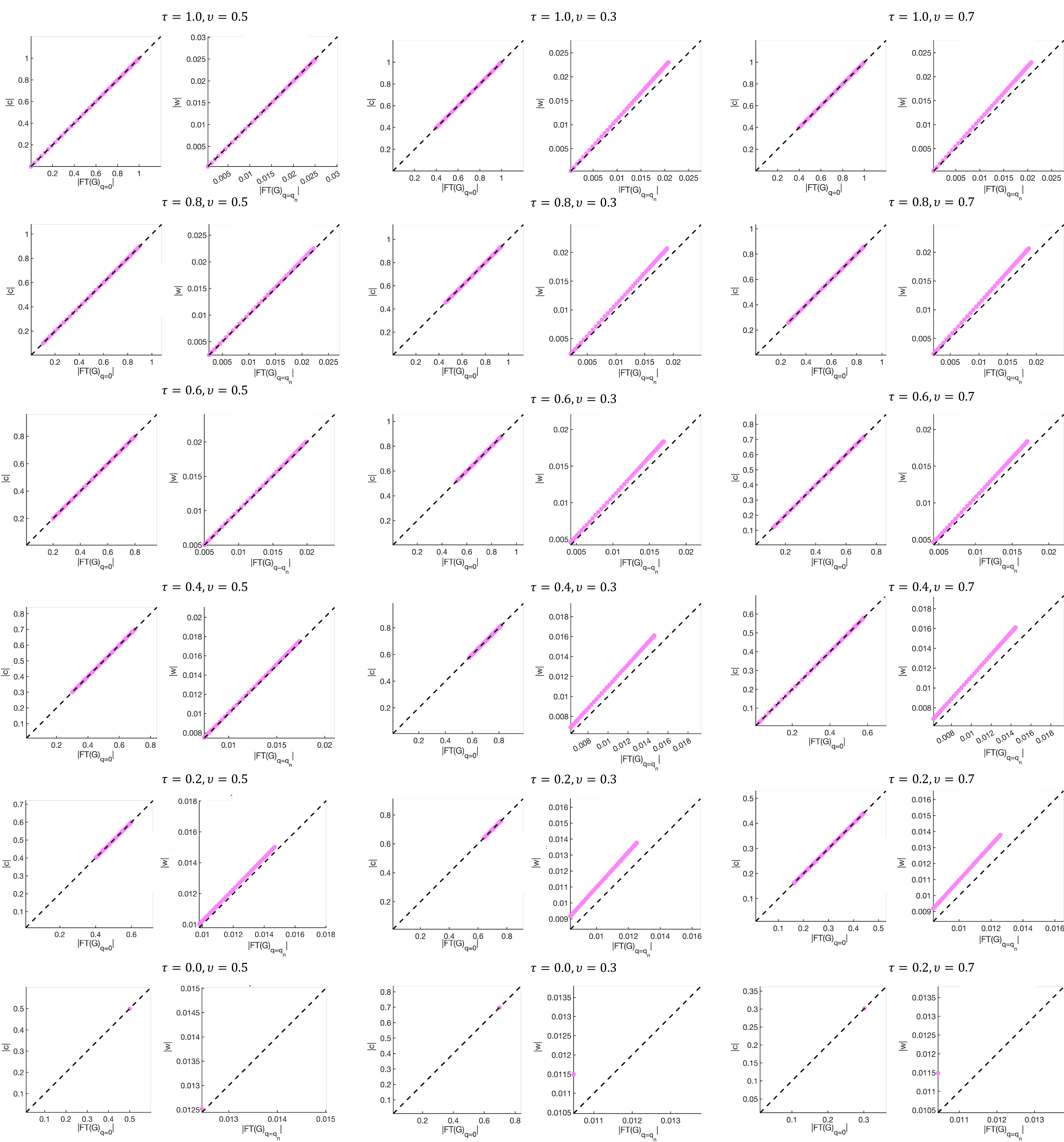}}
\caption{\textbf{Numerical evaluation of spinodal spectral approximation.} Twenty binary spinodal optics with $N=399$ and phase shifts $\delta \Phi$ ranging from $-\pi$ to $\pi$ were generated for the transmission values $\tau$ and duty cycles $\nu$ shown in the plots. For each combination, Fourier coefficients at the zero and characteristic frequencies were calculated both analytically and through equations \ref{eq:w} and \ref{eq:c}. Overall, a high agreement between the two sets is observed which validates the approximation of the spectral representation of binary spinodal optics.}
\label{fig:SI_numerics}
\end{figure*}

\noindent To further corroborate the assumption that the spectrum of a binary spinodal optical element $G(\boldsymbol{r})$ can be approximated by equation \ref{eq:grat}, we performed several numerical calculations for different cases of phase shifts $\delta \Phi$, duty cycles $\nu$, and transmission values $\tau$. We calculate the values of $|c|$ and $|w|$ from equations \ref{eq:w} and \ref{eq:c} and compare them with the numerically calculated mean values of $|\hat{G}(\boldsymbol{q_{n}})|$ and $|\hat{G}(0)|$ (Fig. S\ref{fig:SI_numerics}). As expected, there is a perfect match between $|c|$ and $|\hat{G}(0)|$. For $|w|$, there is an overall high agreement with small discrepancies appearing for duty cycles that severely deviate from 0.5 and for decreasing transmission. In these cases, equation \ref{eq:w} tends to overestimate the actual Fourier coefficients. This can be understood as power leaking away from the well defined ring into a background term. When propagated, this background term will induce discrepancies to the location of the first self image as described in the next section. 

\section{Numerical evaluation of visibility through $M(z)$}

\begin{figure*}[h!] 
\centering
{\includegraphics[width=1\textwidth]{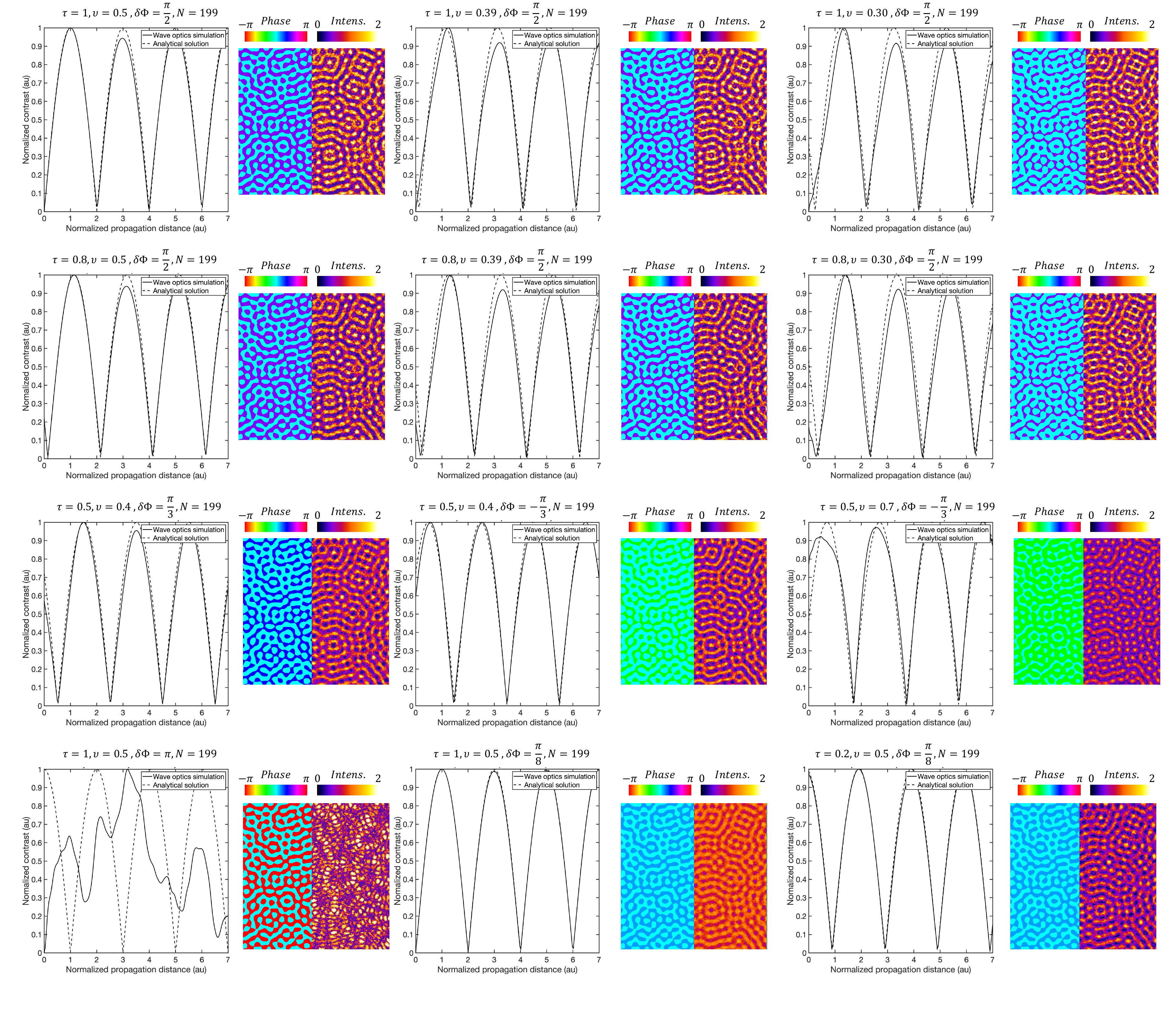}}
\caption{\textbf{Comparison of simulated and theoretically modeled visibilities.} For each panel, normalized visibility (by its maximum value) (continuous curve) calculated from the Fourier transform of the propagated intensity image is compared withe the analytical solution given by the cosine term of equation \ref{eq:modulation} (dashed line). For each simulation the phase of the binary optical element and the simulated intensity at the first maximum are shown in a split view. }
\label{fig:SI_simulation_theory}
\end{figure*}

\noindent To determine the robustness of the assumptions regarding the formulation of the spinodal optics in the frequency space, we present several examples comparing the analytically calculated visibility from the cosine term of equation~\ref{eq:Mz} with the normalized visibility from conventional wave optics simulations which numerically implement the Fresnel propagator. Overall, a high agreement is observed (Fig. S\ref{fig:SI_simulation_theory}) with noticable variations only arising at extreme cases of duty cycles $\nu$ and transmission values $\tau$. In these cases the discrepancies are observed for low propagation distances before the first self image. This is due to power leaking from the well defined ring in the frequency spectrum of spinodal optics. Interestingly, for higher propagation distances this effect is diminished, meaning that any additional term is decaying rather rapidly. Finally, for $\delta \Phi = \pi,~ \tau = 1,~\text{and }\nu = 0.5$ no self image is observed since $|c| = 0$. 

\begin{figure*}[h!] 
\centering
{\includegraphics[width=1\textwidth]{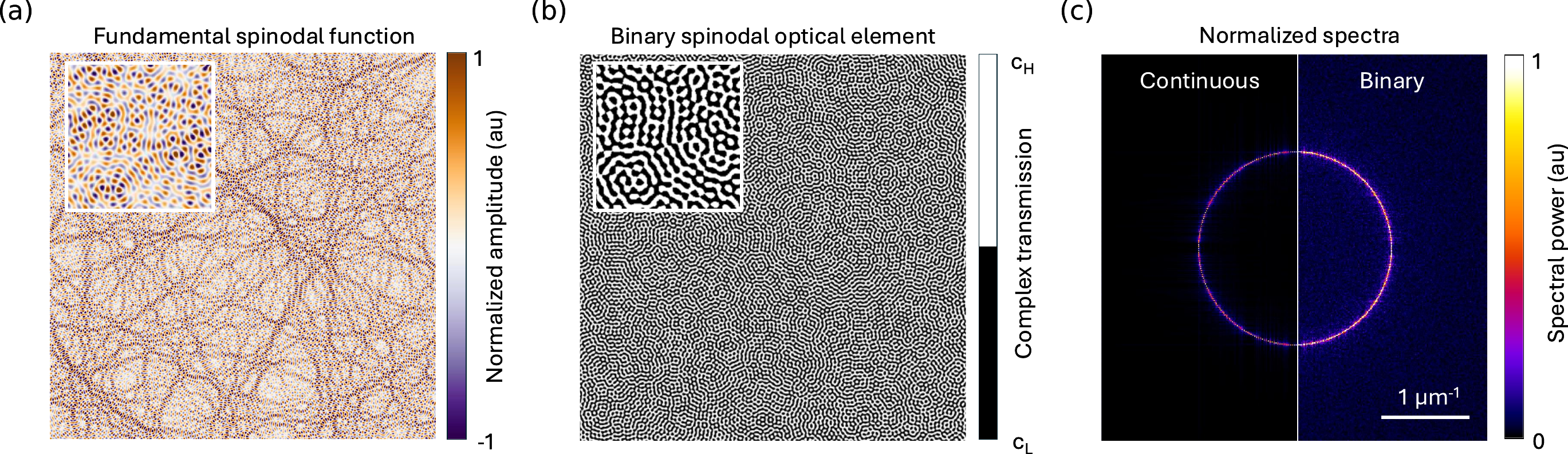}}
\caption{\textbf{Spectral properties of spinodal optics.} (a) Fundamental function of spinal optical element. (b) Binary spinodal optical element. (c) Comparison between normalized spectrum of fundamental function and binary optical element.}
\label{fig:SI_spectral_comparison}
\end{figure*}

\begin{figure*}[h!] 
\centering
{\includegraphics[width=0.7\textwidth]{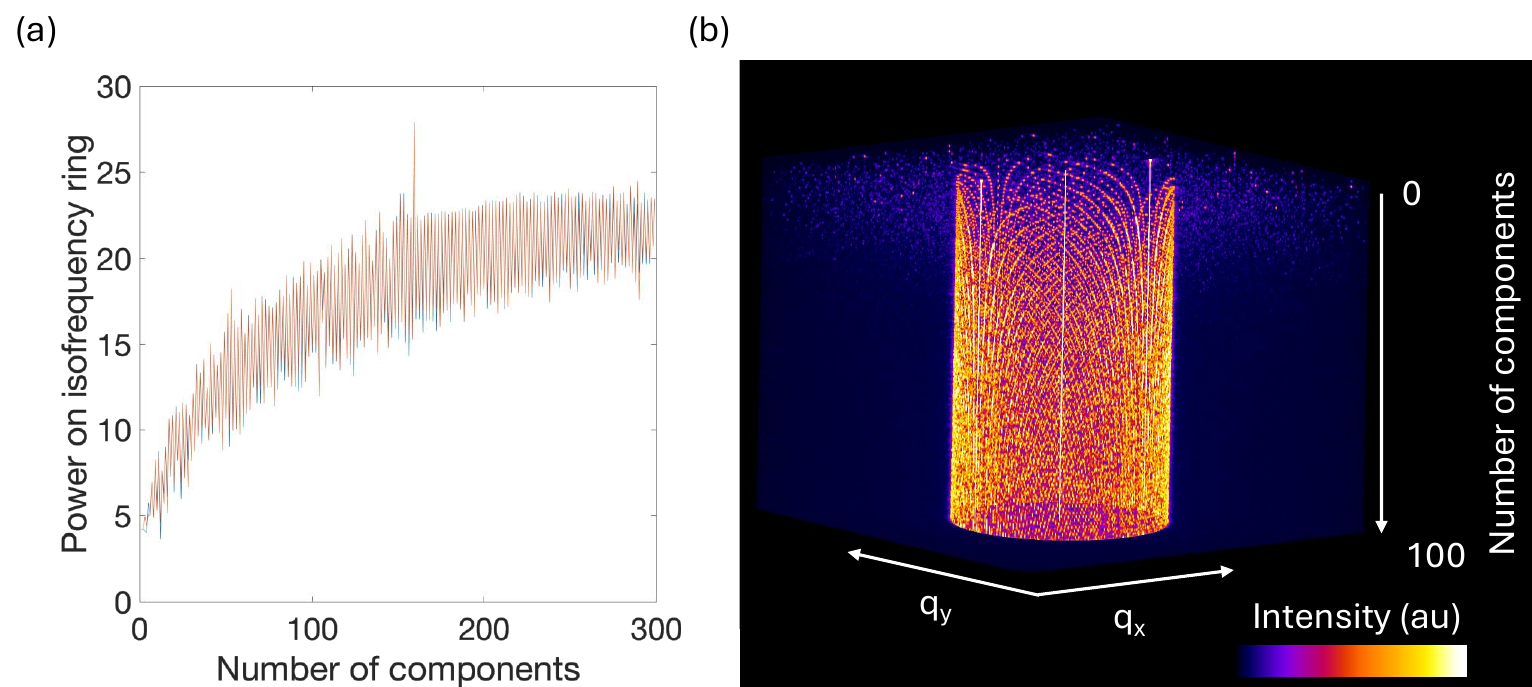}}
\caption{\textbf{Spectral approximation.} (a) For a high number of components $N$ most of the power is distributed on the iso-frequency ring of the modulator. (b) As the number of components increases the background signal reduces in intensity and the spectrum of the modulator can be approximated by a single ring.}
\label{fig:SI_spectral_approximation}
\end{figure*}

\begin{figure*}[h!] 
\centering
{\includegraphics[width=1\textwidth]{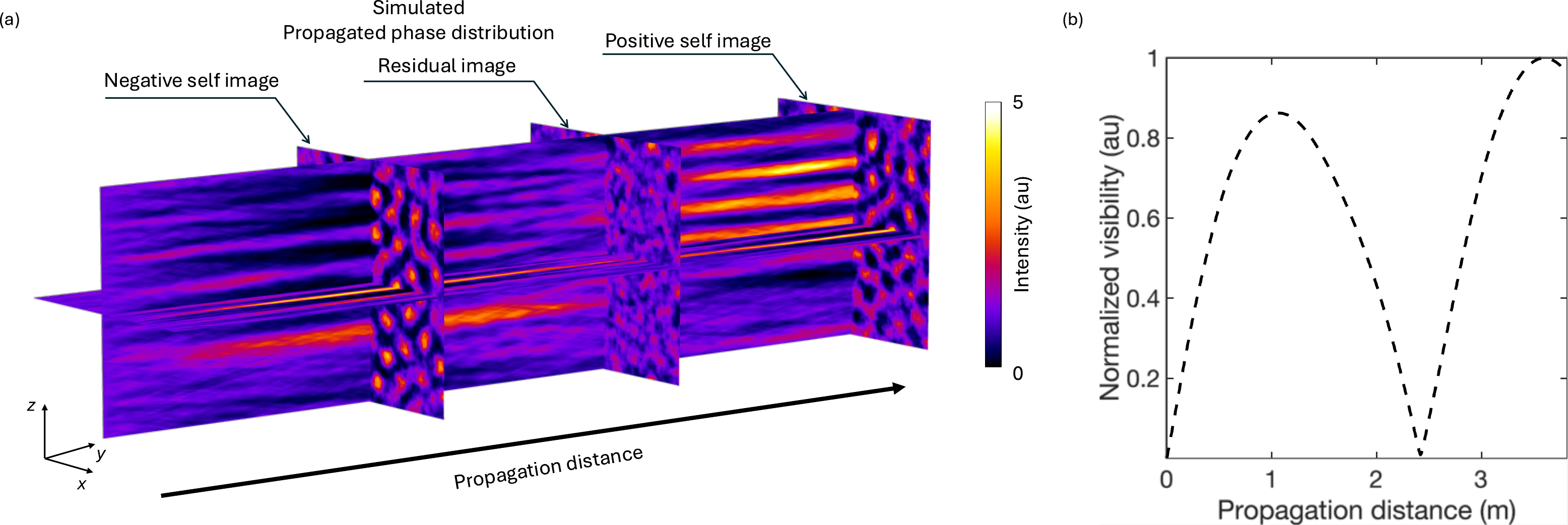}}
\caption{\textbf{Wave optics simulation.} (a) Simulated 3D light distribution through spinodal optics showing the appearance of negative and positive self images. (b) Normalized visibility from simulated patterns. Please note that the visibility at the third position is higher that in the first, as corroborated by experimental data as well.}
\label{fig:SI_simulation}
\end{figure*}

\begin{figure*}[h!] 
\centering
{\includegraphics[width=0.7\textwidth]{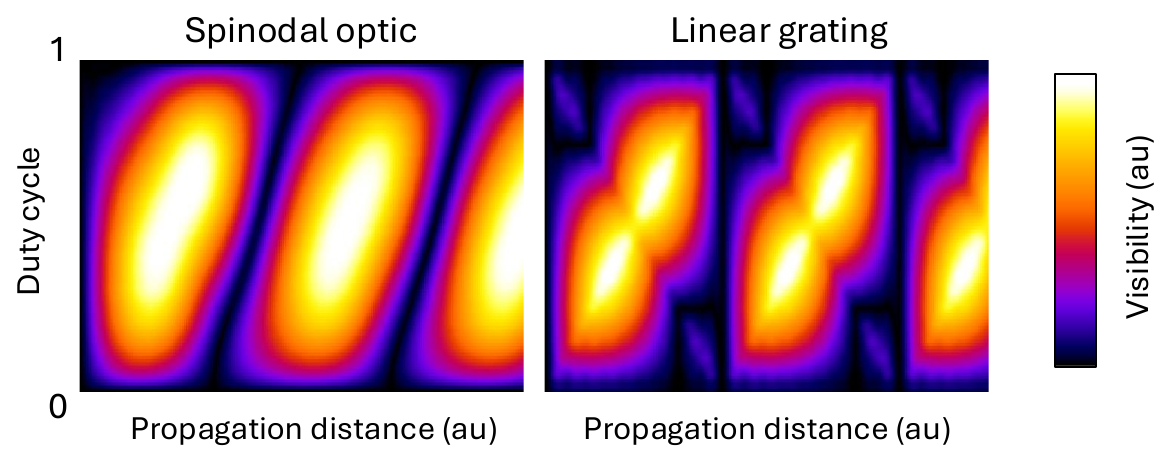}}
\caption{\textbf{Comparison between spinodal and linear optics.} Visibility map comparison between spinodal and linear gratings for varying duty cycle. For the spinodal optic, the visibility maximum continuously covers the propagation space, in contrast the linear gratings shows clear distances that no duty cycle can produce interference pattern.}
\label{fig:SI_linear_comparison}
\end{figure*}

\begin{figure*}[h!] 
\centering
{\includegraphics[width=0.5\textwidth]{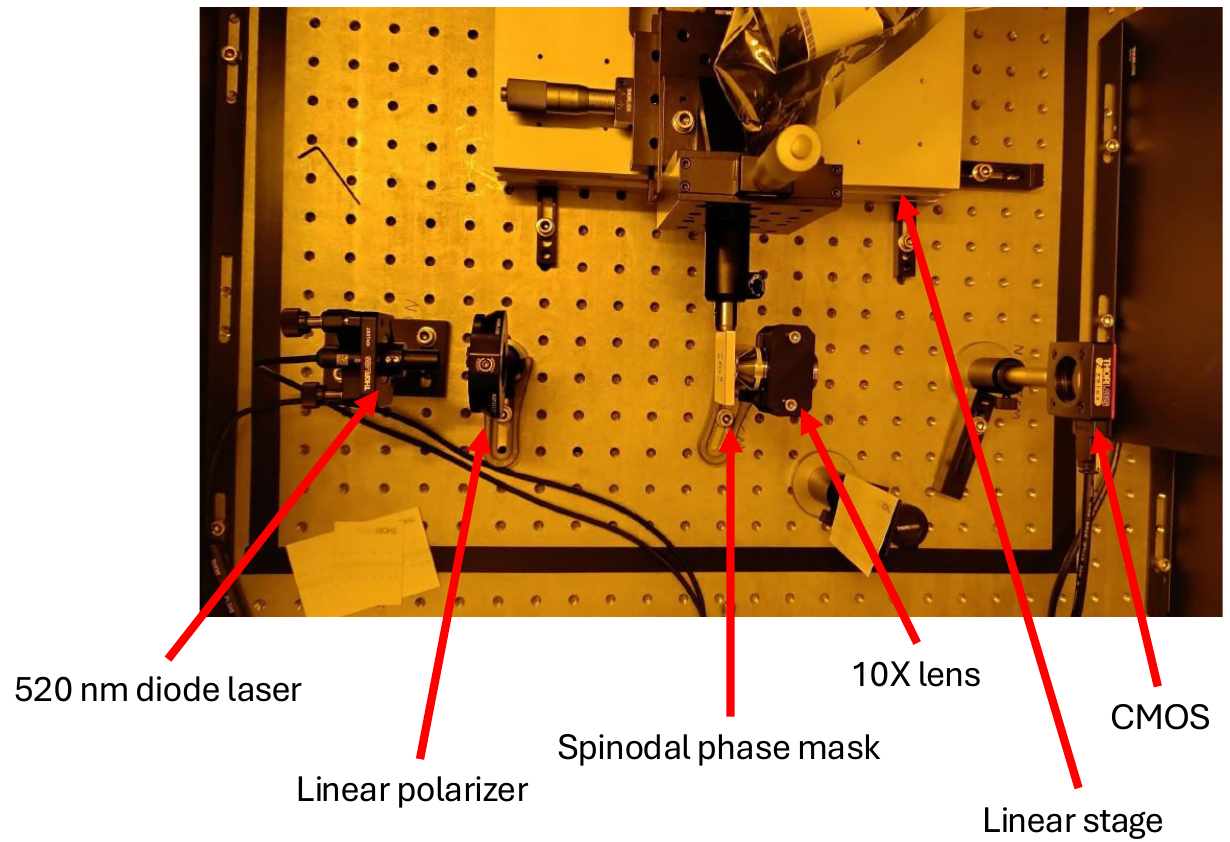}}
\caption{\textbf{Optical experimental setup for demonstration of spinodal Talbot effect.} The beam intensity of a 520~nm diode laser is regulated using a linear polarization filter before reaching the spinodal phase mask. The phase mask is mounted on a linear piezo stage parallel to the optical axis enabling the measurement of different propagation distances $z$. A 10$\times$ magnifying microscope lens and a CMOS camera are mounted fixed on the optical table.}
\label{fig:SI_setup_optical}
\end{figure*}

\begin{figure*}[h!] 
\centering
{\includegraphics[width=1\textwidth]{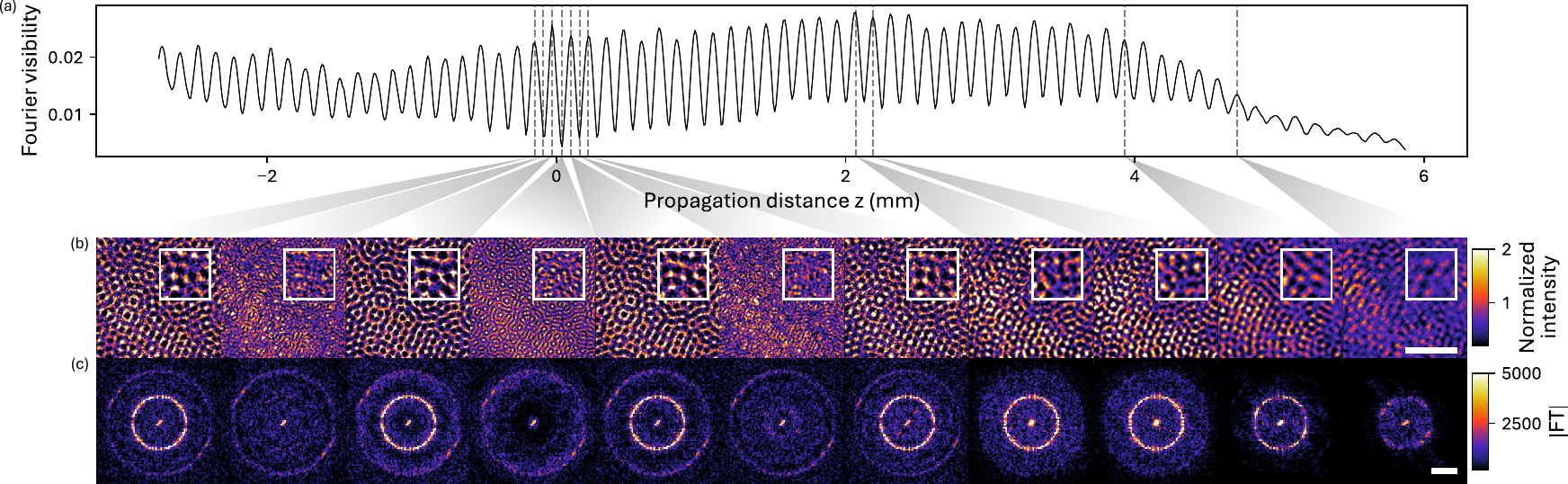}}
\caption{\textbf{Optical propagation measurement.} The results of a propagation measurement in the optical regime (520 nm) for a modulator with characteristic length scale 8 \textmu m are shown. (a) A clear modulation in the Fourier visibility (at the characteristic length scale) for varying propagation distances $z$ can be observed. (b) Flat-field corrected images at selected propagation distances are shown (scale bar 50 \textmu m) together with (c) their Fourier spectra (scale bar 1.0 \textmu m$^{-1}$). The modulation spans over a large number of oscillations. Negative propagation distances $z$ mean a position of the focal plane upstream of the modulator.}
\label{fig:SI_optical}
\end{figure*}

\begin{figure*}[h!] 
\centering
{\includegraphics[width=.8\textwidth]{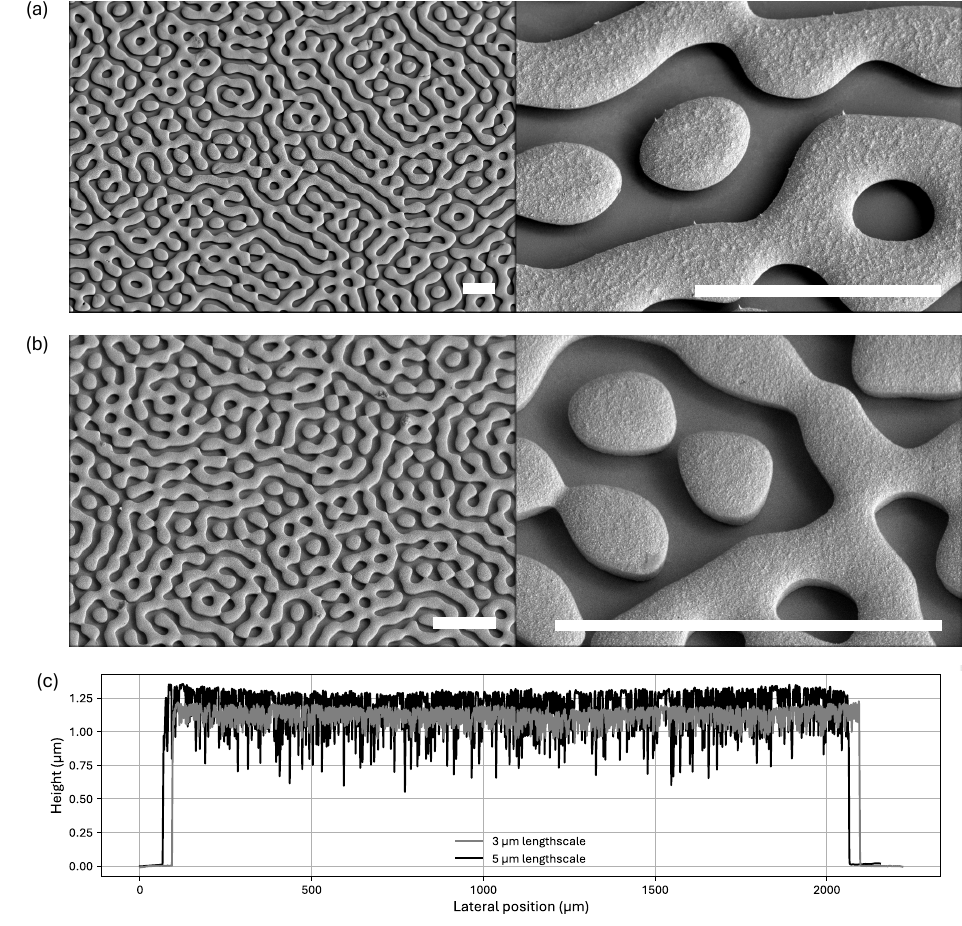}}
\caption{\textbf{Fabricated spinodal optics.} SEM images of the spinodal modulator with characteristic length (a) 5~\textmu m and (b) 3~\textmu m (scale bar 10~\textmu m) are shown. (c) Height profiles along both modulators (acquired with a Bruker XT Stylus Profilometer, 2 \textmu m stylus diameter) are shown. The aim for this optics is to be close to a $\pi/2$ phase shift at $12.8$ keV which corresponds to a gold thickness of $1.37$ \textmu m.}
\label{fig:SI_fabrication}
\end{figure*}

\begin{figure*}[h!] 
\centering
{\includegraphics[width=.8\textwidth]{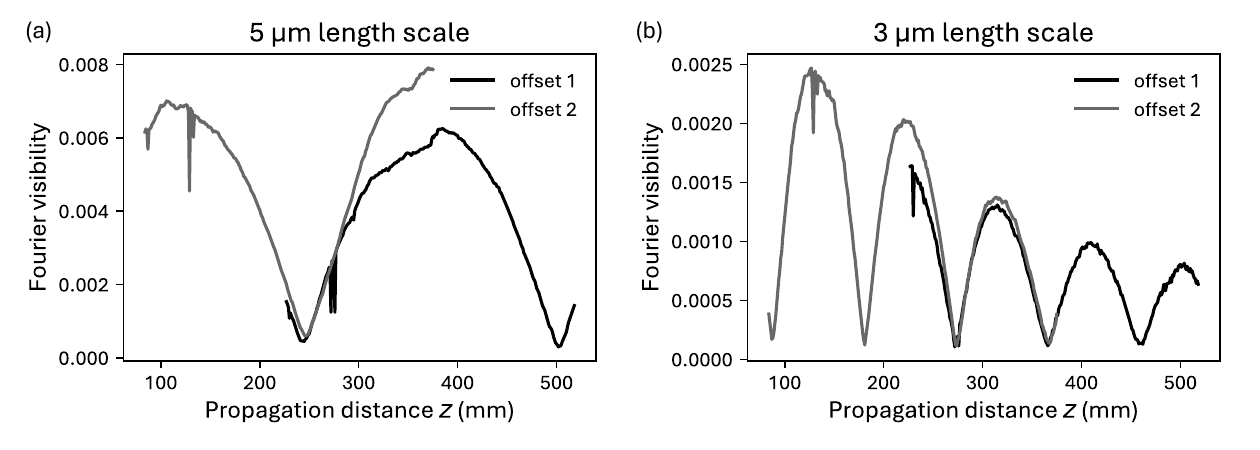}}
\caption{\textbf{X-ray propagation measurement.} The extracted Fourier visibility at the corresponding characteristic length scale is shown for propagation measurements of the 5 \textmu m and 3 \textmu m modulator. The measurement for each modulator consists of two distance scans with different offsets due to the limited range of the utilized stage. The values here are shown prior any normalization of propagation distance or visibility value.}
\label{fig:SI_absolute_visibility}
\end{figure*} 

\begin{figure*}[h!] 
\centering
{\includegraphics[width=.8\textwidth]{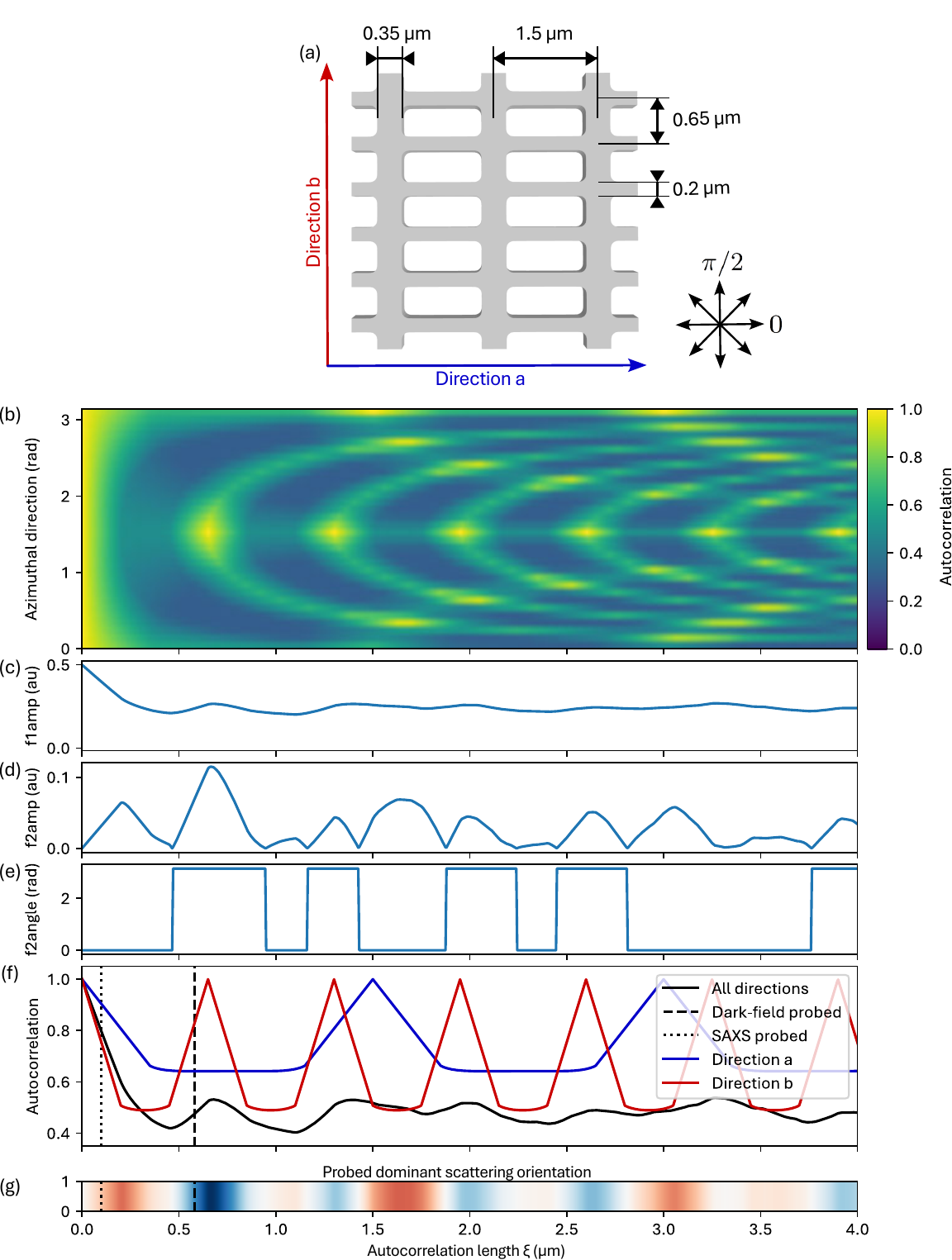}}
\caption{\textbf{Scattering model of the Lepidoptera.} (a) The orientation in the scattering signal of the Lepidoptera sample is modeled using a simple approximation of the scale structure consisting of ribs (vertical) and crossribs (horizontal). The chosen sizes are selected from a series of SEM images. The autocorrelation function of this model is calculated and its (b) polar transformation is shown. The azimuthal angle 0~rad corresponds to the horizontal direction $a$ and the angle $\pi/2$~rad to the vertical direction $b$. The polar representation of the autocorrelation function is Fourier transformed along the azimuthal direction where (c) the amplitude of the first component (DC component) represents the total scattering. (d) The amplitude of the second component represents the degree of orientation and (e) the phase of the second component the dominant scattering direction. Here a phase value of $\pi$~rad corresponds to dominant scattering in direction $a$ and a phase value of $0$~rad corresponds to dominant scattering in direction $b$. (f) The scattering components in the two main directions are shown together with the total scattering signal. (g) Based on the Fourier analysis the dominant scattering direction is visualized where the color represents the direction (f2angle) and the saturation the degree of orientation (f2amp). The model illustrates well the different probed dominant scattering orientations for dark-field and SAXS.}
\label{fig:SI_model}
\end{figure*}

\begin{figure*}[h!] 
\centering
{\includegraphics[width=.7\textwidth]{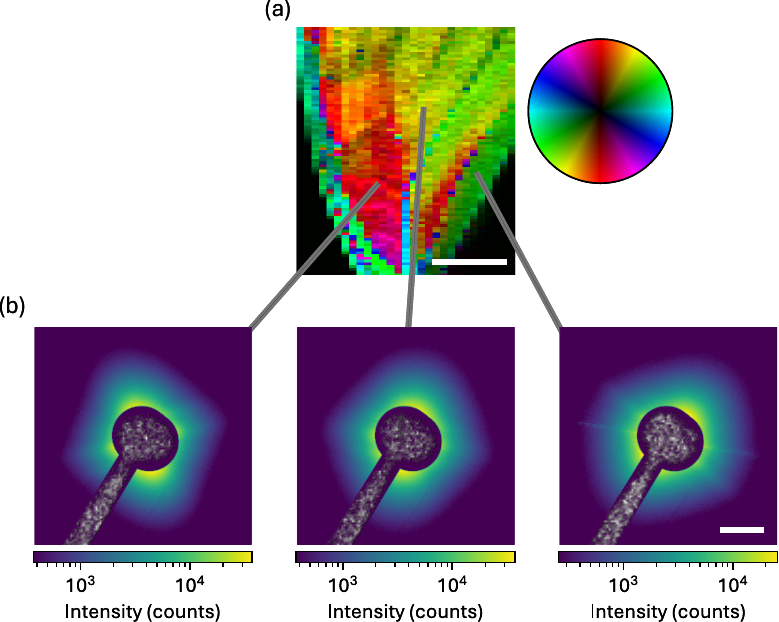}}
\caption{\textbf{SAXS pattern for Lepidoptera.} (a) An orientation SAXS map ($q=0.0064\,\text{\r{A}}^{-1}$, scale bar 5 mm) is shown to illustrate the location of (b) three SAXS scattering pattern (scale bar 50 px respectively 3.75 mm). Only a small part around the beam stop of the SAXS pattern is shown. Other parts of the detector (3.015 m downstream of the sample) have not detected any relevant amount of photons. The $q$-value for the shown scattering pattern is chosen in a way that the extracted signal is as close to the beam-stop as possible while still being azimuthal complete. Each scattering pattern shows two stronger orientations which resemble the symmetry of the structure of the scales. Only the dominant orientation of those is shown in the colored images for the scattering orientation.}
\label{fig:SI_saxs_butterfly}
\end{figure*}

\begin{figure*}[h!] 
\centering
{\includegraphics[width=0.7\textwidth]{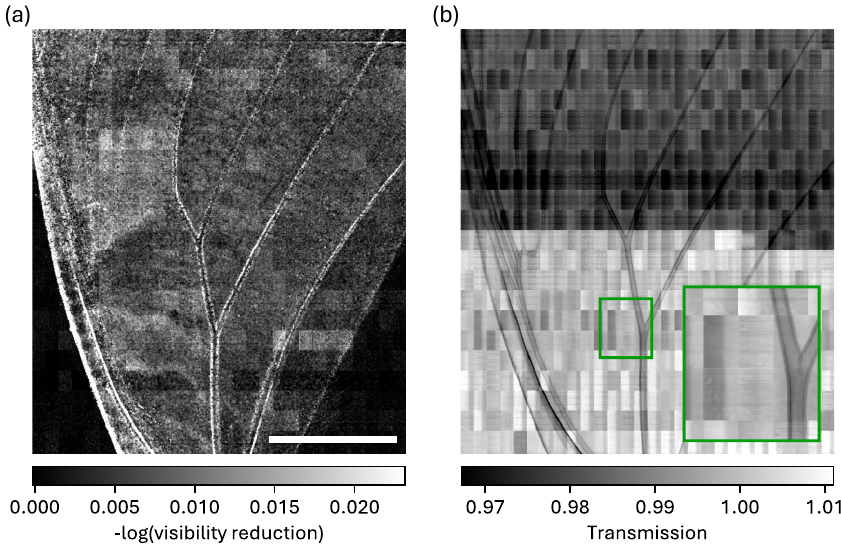}}
\caption{\textbf{Transmission signal of Lepidoptera wing.} The retrieved contrast modalities (a) dark-field and (b) transmission are shown next to each other for comparison (scale bar 5 mm). The transmission image shows the tiles of the stitching process. Only the veins show contrast in the transmission signal but not the scales. The change in transmission in the center of the image is due to top-up in the storage ring.}
\label{fig:SI_butterfly_transmission}
\end{figure*}

\begin{figure*}[h!] 
\centering
{\includegraphics[width=1\textwidth]{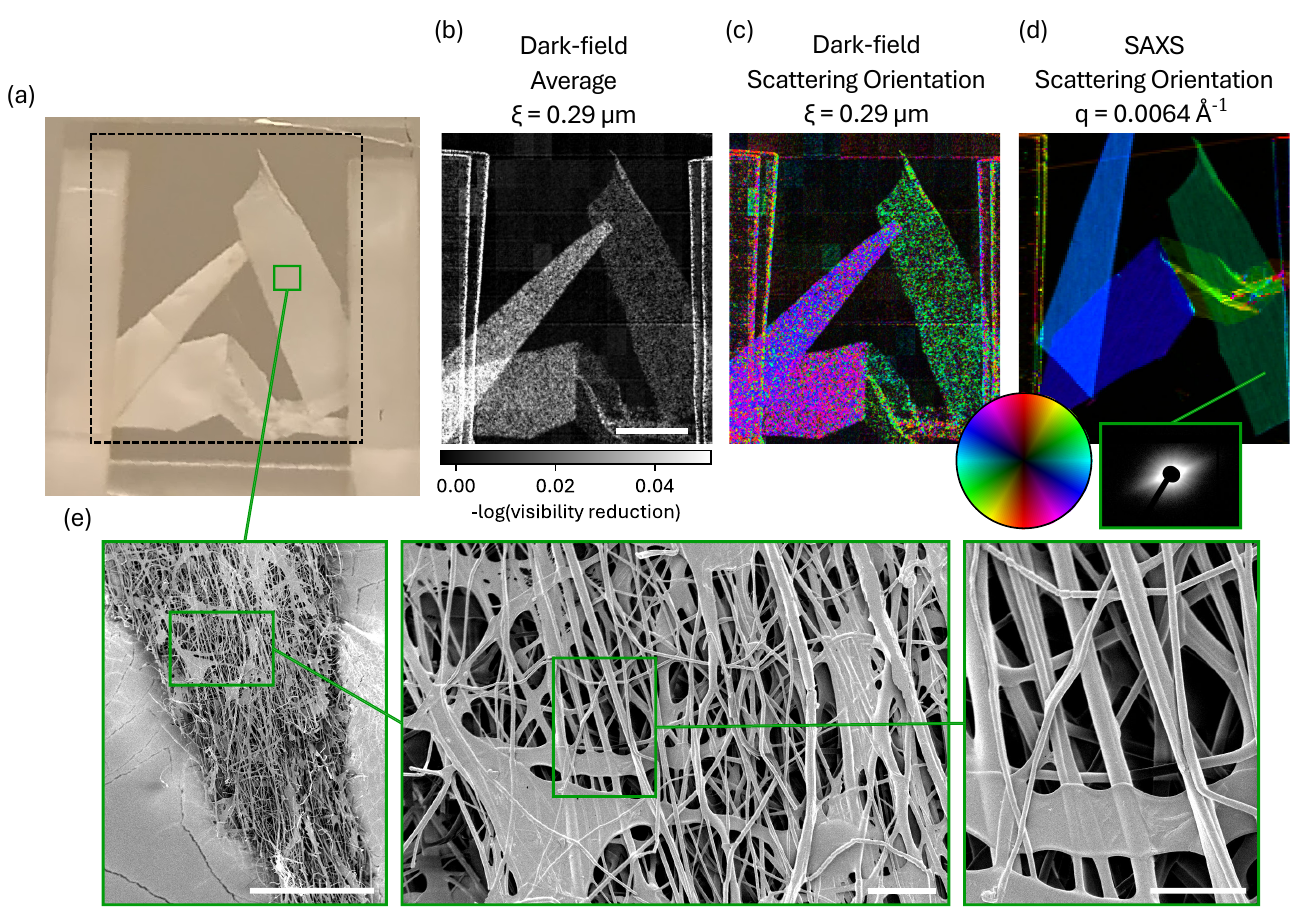}}
\caption{\textbf{Imaging of fibroin silk fibers.} (a) Photo of the mounted fibroin silk fibers organized in sheets. (b) Azimuthal averaged dark-field signal. (scale bar 2 mm) (c) Scattering orientation of the extracted dark-field information. The color encodes the scattering direction and the brightness the azimuthal averaged dark-field signal. (d) Scattering orientation of the SAXS data for $q=0.0064\,\text{\r{A}}^{-1}$. Both dark-field and SAXS signal are strongly orientated with a maximum scattering signal perpendicular to the fibers. A scattering image around the beam stop is shown for one example position of the SAXS map (log scale). (e) Scanning electron micrograph of a fibroin fibers using different magnifications. (scale bars left to right 100~\textmu m, 10~\textmu m, 5~\textmu m)}
\label{fig:SI_fibroin}
\end{figure*}

\begin{figure*}[h!] 
\centering
\includegraphics[width=0.7\textwidth]{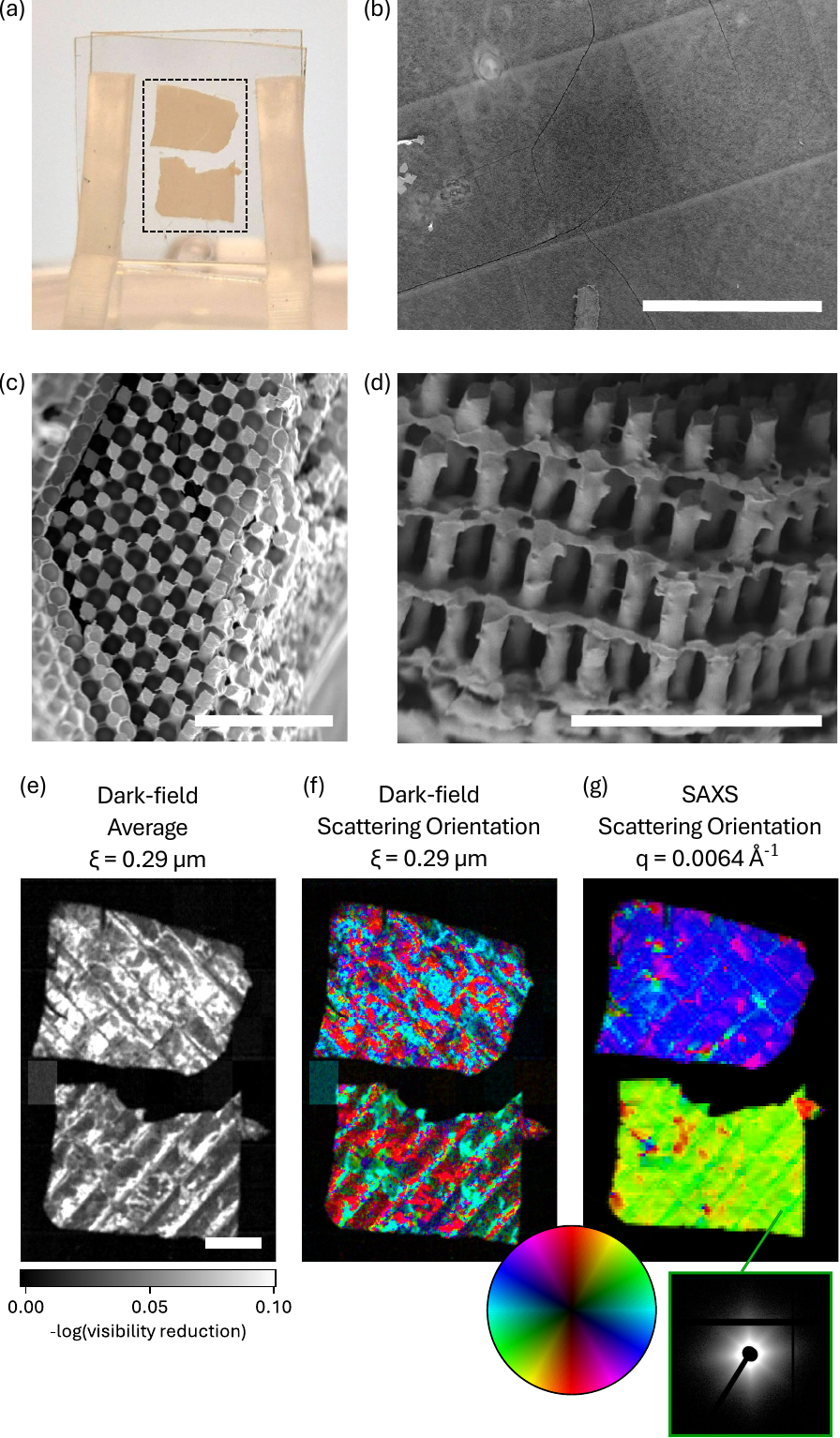}
\caption{\textbf{Imaging of nanoarchitected metamaterials.} Macroscopic images of a nanoarchitected metamaterial showing (a) a photograph and (b) SEM image of the surface (scale bar 0.5 mm). SEM images showing an (c) intermediate layer and (d) cross section (scale bars 5 \textmu m). X-ray images showing the dark-field signal collected with spinodal optics (e \& f) and (g) SAXS data (scale bar 1 mm). The color encodes the dominant scattering direction and the brightness the azimuthal averaged scattering strength. A scattering image around the beam stop is shown for one example position of the SAXS map (log scale).}
\label{fig:SI_architected}
\end{figure*}

\begin{figure*}[h!] 
\centering
{\includegraphics[width=.9\textwidth]{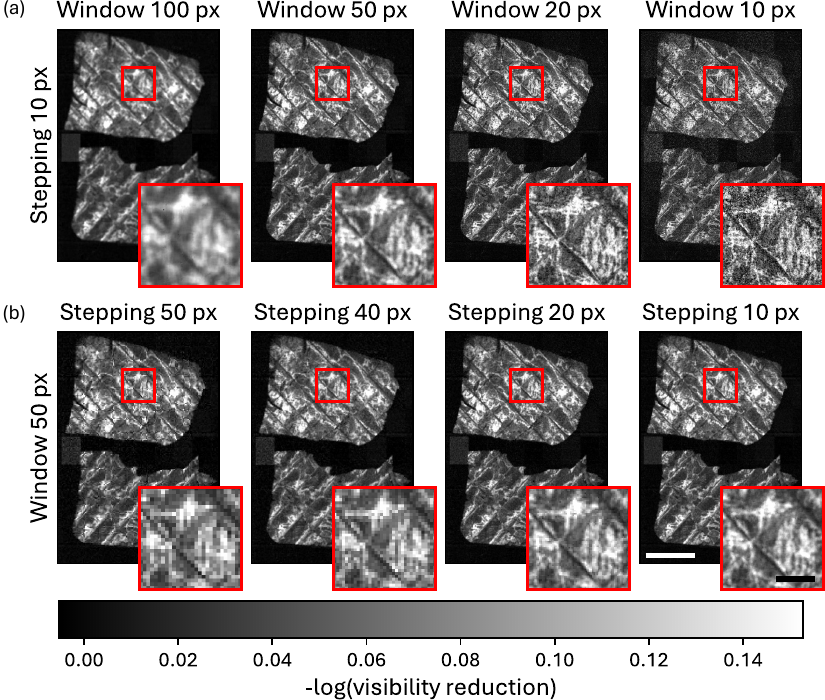}}
\caption{\textbf{Parameters for dark-field retrieval.} (a) The retrieved dark-field signal is shown for different sizes of the analyzer window, while the step width of the window is constant. A decrease in window size shows clearly a gain in resolution which goes for small window sizes on the cost of signal to noise ratio. (b) The retrieved dark-field signal is shown for a constant size of the analyzer window (50 px $\times$ 50 px) but different step sizes (scale bar 2 mm in large image and 0.5 mm in inset). A gain in resolution is visible for stepping the window and having overlapping analyzer windows.}
\label{fig:SI_window_test}
\end{figure*}

\begin{figure*}[h!] 
\centering
{\includegraphics[width=.8\textwidth]{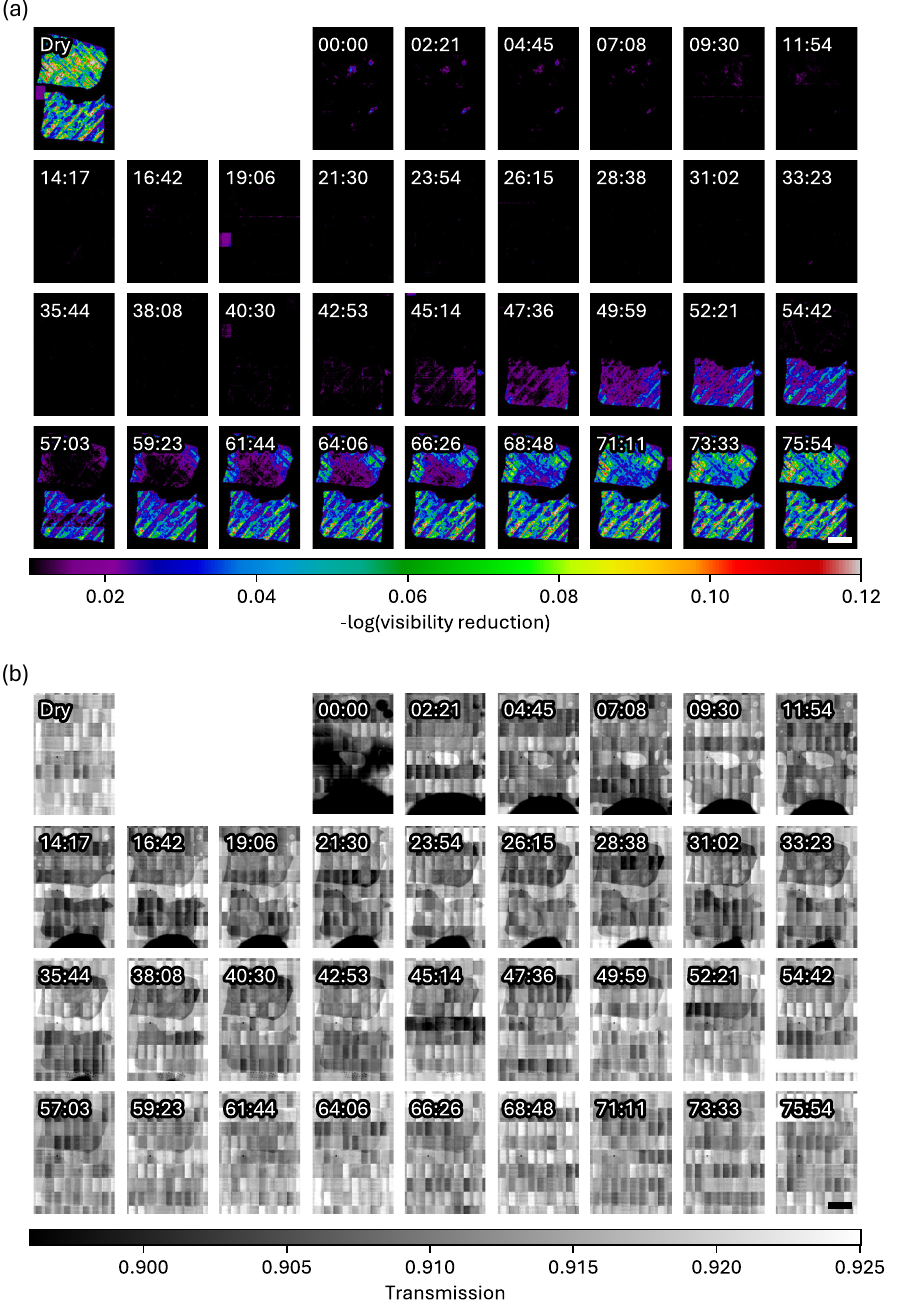}}
\caption{\textbf{Drying of nanoarchitected metamaterial.} The complete drying process of the nanoarchitected metamaterial is shown including (a) extracted dark-field information and (b) transmission signal (scale bar 1~mm, timestamp of the acquisition start for each projection in min:sec). After the start of the experiment the transmission signal shows a droplet on the bottom of the sample which is not visible in the dark-field signal.}
\label{fig:SI_timeseries}
\end{figure*}

\end{document}